\documentclass[aps,pra]{revtex4-2}
\usepackage{amsfonts}
\usepackage{amsmath}
\usepackage{amsthm}
\usepackage{amssymb}
\usepackage{graphicx}
\usepackage{appendix}
\usepackage{hyperref}
\usepackage{textcomp}
\usepackage{multirow}
\usepackage{color}
\usepackage{bm}
\usepackage{braket}
\usepackage{subfigure}
\begin{document}

\title{Discrete and semi-discrete multidimensional solitons and vortices: Established results and novel findings}
\author{Boris A. Malomed}
\affiliation{Department of Physical Electronics, School of Electrical Engineering, Tel Aviv University, Tel Aviv 69978, Israel\\
Instituto de Alta Investigaci\'{o}n, Universidad de Tarapac\'{a},
Casilla 7D, Arica, Chile}

\begin{abstract}
This article presents a concise survey of basic discrete and semi-discrete
nonlinear models which produce two- and three-dimensional (2D and 3D)
solitons, and a summary of main theoretical and experimental results
obtained for such solitons. The models are based on the discrete nonlinear
Schr{\"{o}}dinger (DNLS) equations and their generalizations, such as a
system of discrete Gross-Pitaevskii (GP) equations with the Lee-Huang-Yang
corrections, the 2D Salerno model (SM), DNLS equations with long-range
dipole-dipole and quadrupole-quadrupole interactions, a system of coupled
discrete equations for the second-harmonic generation with the quadratic ($%
\chi ^{(2)}$) nonlinearity, a 2D DNLS equation with a superlattice
modulation opening mini-gaps, a discretized NLS equation with rotation, a
DNLS coupler and its $\mathcal{PT}$-symmetric version, a system of DNLS
equations for the spin-orbit-coupled (SOC) binary Bose-Einstein condensate,
and others. The article presents a review of basic species of
multidimensional discrete modes, including fundamental (zero-vorticity) and vortex solitons,
their bound states, gap solitons populating mini-gaps, symmetric and
asymmetric solitons in the conservative and $\mathcal{PT}$-symmetric
couplers, cuspons in the 2D SM, discrete SOC solitons of the semi-vortex and
mixed-mode types, 3D discrete skyrmions, and some others.
\end{abstract}

\maketitle

\section{Introduction}

\subsection{Discrete nonlinear Schr\"{o}dinger (DNLS) equations}

\subsubsection{The basic equation}

Commonly adopted models of physical media are based on linear and nonlinear
partial differential equations, such as the Gross-Pitaevskii (GP) equations
for the mean-field wave function $\psi \left( x,y,z;t\right) $ in atomic
Bose-Einstein condensates (BECs) \cite{Pit}, and nonlinear Schr\"{o}dinger
(NLS) equations for the amplitude of the electromagnetic field in optical
waveguides \cite{KA,Gadi}. The scaled form of the three-dimensional (3D)
GP/NLS equation is%
\begin{equation}
i\psi _{t}=-(1/2)\nabla ^{2}\psi +\sigma |\psi |^{2}\psi +U\left(
x,y,z\right) \psi ,  \label{NLSE}
\end{equation}%
where $\sigma =+1$ and $-1$ correspond to the self-defocusing and focusing
signs of the cubic nonlinearity, and $U\left( x,y,z\right) $ is an external
potential. In the application to optics, time $t$ is replaced, as the
evolution variable, by the propagation distance $z$, while the original
coordinate $z$ is then replaced by the temporal one, $\tau =t-z/V_{\mathrm{gr%
}}$, where $V_{\mathrm{gr}}$\ is the group velocity of the carrier wave \cite%
{KA}. In optics, the effective potential may be two-dimensional (2D), $%
-U(x,y)$, which represents a local variation of the refractive index in the
transverse plane.

In many cases the potential is spatially periodic, such as one induced by
optical lattices (OLs) in BEC \cite{OL,Porter}, or by photonic crystals
which steer the propagation of light in optics \cite{PhotCryst},
\begin{equation}
U_{\mathrm{latt}}\left( x,y,z\right) =-\varepsilon \left[ \cos (2\pi
L/x)+\cos (2\pi L/y)+\cos (2\pi L/z)\right] ,  \label{UOL}
\end{equation}%
as well as its 2D and 1D reductions. A deep lattice potential, which
corresponds to large amplitude $\varepsilon $ in Eq. (\ref{UOL}), splits the
continuous wave function into a set of \textquotedblleft
droplets\textquotedblright\ trapped in local potential wells, which are
linearly coupled by tunneling. Accordingly, in the framework of the \textit{%
tight-binding approximation}, the NLS equation is replaced by a discrete NLS
(DNLS) equation with the linear coupling between adjacent sites of the
discrete lattice (nearest neighbors). This equation was derived, in the 1D
form, for arrays of optical fibers \cite%
{ChristoJoseph,Silberberg,discr-review-0,discr-review} and plasmonic
nanowires \cite{Nicolae}, as well as for BEC loaded in a deep OL potential
\cite{Smerzi}:%
\begin{gather}
i\dot{\psi}_{l,m,n}=-(1/2)\left[ \left( \psi_{l+1,m,n}+\psi _{l-1,m,n}-2\psi
_{l,m,n}\right) +\left( \psi _{l,m+1,n}+\psi _{l,m-1,n}-2\psi
_{l,m,n}\right) \right.   \notag \\
\left. +\left( \psi _{l,m,n+1}+\psi _{l,m,n-1}-2\psi _{l,m,n}\right) \right]
+\sigma \left\vert \psi _{l,m,n}\right\vert ^{2}\psi _{l,m,n}+V_{l,m,n}\psi
_{l,m,n},  \label{DNLSE}
\end{gather}%
where the overdot stands for $d/dt$, and the set of integer indices, $\left(
l,m,n\right) $, replaces continuous coordinates $\left( x,y,z\right) $ in
Eq. (\ref{NLSE}). In Eq. (\ref{DNLSE}), potential $V_{l,m,n}$ is a possible
smooth addition to the deep lattice potential which imposes the
discretization.

The rigorous mathematical derivation of the DNLS equation by the
discretization of the underlying continuum NLS equation with the deep
spatially periodic potential is based on the expansion of the continuous
wave field over the set of Wannier functions \cite{Wannier}. These are
linear combinations of the quasiperiodic Bloch wave functions which feature
shapes localized around potential minima \cite{Wannier-review}, thus
offering a natural basis for the transition to the discrete limit.

The full DNLS equation (\ref{DNLSE}) is often reduced to its 2D and 1D
forms. 1D lattices
are sometimes built in the form of zigzag chains, making it relevant to add
couplings between the next-nearest neighbors \cite{NNN,ChongNNN}. 2D
lattices with similar additional couplings were elaborated too \cite{NNN2D}.

Unlike the continuous NLS equation (\ref{NLSE}), the nonlinearity signs $%
\sigma =\pm 1$ in Eq. (\ref{DNLSE}) are actually equivalent to each other.
Indeed, the \textit{staggering transformation} of the discrete wave function,%
\begin{equation}
\psi _{l,m,n}(t)\equiv (-1)^{l+m+n}\exp \left( -6it\right) \tilde{\psi}%
_{l,m,n}^{\ast }(t)  \label{stagg}
\end{equation}%
where $\ast $ stands for the complex-conjugate, transforms Eq. (\ref{DNLSE})
with $\sigma =-1$ into the same equation for $\tilde{\psi}_{l,m,n}$ with $%
\sigma =+1$. For the 2D and 1D DNLS equations, $\exp \left( -6it\right) $ in
the corresponding versions of staggering transformation (\ref{stagg}) is
replaced, respectively, by $\exp \left( -4it\right) $ and $\exp \left(
-2it\right) $.

\subsubsection{Extended equations}

The DNLS equation and its extensions, such as systems of coupled DNLS
equations \cite{Angelis,Herring}, constitute a class of models with a large
number of physical realizations \cite{Aubry,discr-review,Rothos,Tsoy,chapter}%
. They have also drawn much interest as subjects of mathematical studies
\cite{DNLS-book}. One of incentives for this interest is the fact that the
discreteness arrests the development of the \textit{critical} and \textit{%
supercritical} \textit{collapse}, which is driven by the self-focusing
nonlinear term with $\sigma =-1$ in the 2D and 3D continuous NLS equations (%
\ref{NLSE}), respectively. The collapse leads to the emergence of singular
solutions in the form of infinitely tall peaks, after a finite evolution
time \cite{Gadi}.\ Naturally, the discreteness causes the arrest of the
collapse, replacing it by a \textit{quasi-collapse} \cite{Laedke}, when the
width of the shrinking peak becomes comparable to the spacing of the DNLS
lattice.

The possibility of the collapse destabilizes formal 2D and 3D soliton
solutions which are produced by Eq. (\ref{NLSE}), therefore a challenging
problem is prediction and experimental realization of physical settings than
make it possible to produce stable multidimensional solitons \cite{NRP,book}%
. Thus, the discreteness provides a general method for the stabilization of
2D and 3D solitons.

\paragraph{The Gross-Pitaevskii (GP) equations amended by effects of quantum
fluctuations}

Another promising possibility for the suppression of the collapse is offered
by binary BEC, in which the cubic inter-component attraction creates 3D
soliton-like states in the form of \textquotedblleft quantum droplets"
(QDs), while the development of the supercritical collapse is arrested by
the self-repulsive quartic term that takes into account a correction to the
mean-field nonlinearity produced by quantum fluctuations (known as the
celebrated\textit{\ Lee-Huang-Yang} effect \cite{LHY}). For the symmetric
binary condensate, with identical wave functions $\psi $ of its components,
the accordingly amended scaled GP equations (in the absence of the trapping
potential) was derived by Petrov \cite{Petrov}
\begin{equation}
i\psi _{t}=-(1/2)\nabla ^{2}\psi -|\psi |^{2}\psi +|\psi |^{3}\psi .
\label{amended GP}
\end{equation}%
Surprisingly quickly, the QD modes predicted by Eq. (\ref{amended GP}) have
been created experimentally, in the quasi-2D \cite{Tarruell1,Tarruell2} and
full 3D \cite{Inguscio} forms. The reduction of the spatial dimension to 2D
and 1D replaces Eq. (\ref{amended GP}) by the following GP equations,
respectively \cite{Petrov-Astra}:%
\begin{eqnarray}
\text{2D}\text{: \ } &&i\psi _{t}=-(1/2)\nabla ^{2}\psi +|\psi |^{2}\ln
(|\psi |^{2})\cdot \psi ,  \label{amended 2D} \\
\text{1D}\text{: \ } &&i\psi _{t}=-(1/2)\psi _{xx}+|\psi |^{2}\psi -|\psi
|\psi .  \label{amended 1D}
\end{eqnarray}%
Note, in particular, that in the 1D equation (\ref{amended 1D}) the quantum
correction is represented by the \emph{attractive} term $-|\psi |\psi $, on
the contrary to its repulsive counterpart $+|\psi |^{3}\psi $ in the 3D
equation (\ref{amended GP}). For this reason, the usual mean-field cubic
term is taken in Eq. (\ref{amended 1D}) with the self-repulsion sign, to
make it possible to study effects of competition of the quadratic
self-attraction and cubic repulsion \cite{Grisha}. A semi-discrete version
of Eq. (\ref{amended 1D}) is considered below, see Eq. (\ref{GPE array}).

\paragraph{The Ablowitz-Ladik (AL) and Salerno-model (SM) equations}

The 1D continuous NLS equation without an external potential is integrable
by means of the inverse-scattering transform, with either sign of the
nonlinearity, $\sigma =\pm 1$ \cite{Zakh,Segur,Calogero,Newell}, although 2D
and 3D extensions of the NLS equation are nonintegrable. The straightforward
discretization destroys the integrability of the 1D NLS equation \cite%
{Herbst,non-integrable}. Nevertheless, the NLS equation admits a specially
designed 1D discretization, which leads to an integrable discrete model,
\textit{viz}., the Ablowitz-Ladik (AL) equation \cite{AL}:
\begin{equation}
i\dot{\psi}_{n}=-\left( \psi _{n+1}+\psi _{n-1}\right) \left( 1+\mu
\left\vert \psi _{n}\right\vert ^{2}\right) ,  \label{ALproper}
\end{equation}%
where positive and negative values of the nonlinearity coefficient, $\mu $,
correspond to the self-focusing and defocusing, respectively. Integrable
discrete equations, such as the AL one, are exceptional models which provide
exact solutions for discrete solitons \cite{Suris}.

Equation (\ref{ALproper}) gives rise to an exact solution for solitons in
the case of $\mu >0$. Setting $\mu \equiv +1$ by means of rescaling, the
solution is%
\begin{equation}
\psi _{n}(t)=\left( \sinh \beta \right) \mathrm{sech}\left[ \beta (n-\xi (t))%
\right] \exp \left[ i\alpha \left( n-\xi (t)\right) -i\varphi (t)\right] ,
\label{ALsoliton}
\end{equation}%
where $\beta $ and $\alpha $ are arbitrary real parameters that determine
the soliton's amplitude, $A\equiv \sinh \beta $, its velocity, $V\equiv \dot{%
\xi}=2\beta ^{-1}\left( \sinh \beta \right) \sin \alpha $, and overall
frequency\linebreak 
$\Omega \equiv \dot{\varphi}=-2\left[ \left( \cosh \beta \right) \cos \alpha
+(\alpha /\beta )\left( \sinh \beta \right) \sin \alpha \right] $. The
existence of exact solutions for traveling solitons in the discrete system
is a highly nontrivial property of the AL equation, which follows from its
integrability. Generically, motion of a discrete soliton through a lattice
is braked by emission of radiation, even if this effect may seem very weak
in direct simulations \cite{Feddersen}. Another integrable discrete model
which admits exact solutions for moving solitons is the Toda-lattice
equation for real coordinates $x_{n}(t)$ of particles with unit mass and
exponential potential of interaction between adjacent ones \cite{Toda}:%
\begin{equation}
\ddot{x}_{n}+\,\exp \left( -\left( x_{n+1}-x_{n}\right) \right) -\exp \left(
-\left( x_{n}-x_{n-1}\right) \right) =0.  \label{TLEu}
\end{equation}

Considerable interest was also drawn to the nonintegrable combination of the
AL and DNLS equations, in the form of the \textit{Salerno model} (SM) \cite%
{SA}, with an additional onsite cubic term, different from the intersite one
in Eq. (\ref{ALproper}):%
\begin{equation}
i\dot{\psi}_{n}=-\left( \psi _{n+1}+\psi _{n-1}\right) \left( 1+\mu
\left\vert \psi _{n}\right\vert ^{2}\right) -2\left\vert \psi
_{n}\right\vert ^{2}\psi _{n}.  \label{SAmodel}
\end{equation}%
Here, the signs and magnitude of the onsite nonlinearity coefficient are
fixed by means of the staggering transformation (\ref{stagg}) and rescaling.
The SM finds a physical realization in the context of the Bose-Hubbard
model, which represents the BEC loaded in a deep OL, taking into regard the
nonlinearity of the intersite coupling \cite{BH-review}.

The AL and SM equations (\ref{ALproper}) and (\ref{SAmodel}) conserve the
total norm $N$, whose definition is different from the straightforward one
for the DNLS equation, given below by Eq. (\ref{Ndiscr}); namely,
\begin{equation}
N_{\mathrm{AL,SM}}=(1/\mu )\sum_{n}\ln \left\vert 1+\mu |\psi
_{n}|^{2}\right\vert  \label{eq:Norm}
\end{equation}%
\cite{AL,Cai}. The Hamiltonians of the AL and SM equations, which are
dynamical invariants too, are%
\begin{equation}
H_{\mathrm{AL}}=-\sum_{n}\left( \psi _{n}\psi _{n+1}^{\ast }+\psi _{n+1}\psi
_{n}^{\ast }\right) ,  \label{HAL}
\end{equation}%
\begin{equation}
H_{\mathrm{SM}}=-\sum_{n}\left[ \left( \psi _{n}\psi _{n+1}^{\ast }+\psi
_{n+1}\psi _{n}^{\ast }\right) +(2/\mu )|\psi _{n}|^{2}\right] +(2/\mu )N_{%
\mathrm{SM}}.  \label{HSM}
\end{equation}%
In particular, the ostensible \textquotedblleft
simplicity\textquotedblright\ of Hamiltonian (\ref{HAL}) is related to the
complexity of the respective Poisson brackets (symplectic structure), which
determine the evolution equations for $\psi _{n}$ as $d\psi _{n}/dt=\left\{
H,\psi _{n}\right\} $. For the AL and SM models, the Poisson brackets,
written for a pair of arbitrary functions $B\left( \psi _{n},\psi _{n}^{\ast
}\right) $ and $C\left( \psi _{n},\psi _{n}^{\ast }\right) $, are \cite%
{SA,BH-review}
\begin{equation}
\left\{ B,C\right\} =i\sum_{n}\left( \frac{\partial B}{\partial \psi _{n}}%
\frac{\partial C}{\partial \psi _{n}^{\ast }}-\frac{\partial B}{\partial
\psi _{n}^{\ast }}\frac{\partial C}{\partial \psi _{n}}\right) \left( 1+\mu
\left\vert \psi _{n}\right\vert ^{2}\right) .  \label{Poisson}
\end{equation}

It is also relevant to consider the continuum limit of the SM, which is
introduced by approximating the intersite combination of the discrete fields
by a truncated Taylor's expansion,
\begin{equation}
\psi _{n}(t)\equiv e^{2it}\Psi (x,t),\qquad \Psi \left( x=n\pm 1,t\right)
\approx \Psi \left( x=n\right) \pm \Psi _{x}{\big |}_{x=n}+(1/2)\Psi _{xx}{%
\big |}_{x=n}\,,  \label{psiPsi}
\end{equation}%
where $\Psi (x)$ is considered as a function of continuous coordinate $x$,
whose integer values coincide with the discrete lattice coordinate $n$. The
substitution of approximation (\ref{psiPsi}) in Eq. (\ref{SAmodel}) leads to
a generalized (nonintegrable) NLS equation \cite{Zaragoza}
\begin{equation}
i\Psi _{t}=-\left( 1+\mu \left\vert \Psi \right\vert ^{2}\right) \Psi
_{xx}-2\left( 1+\mu \right) \left\vert \Psi \right\vert ^{2}\Psi .
\label{SAcont}
\end{equation}%
Equation (\ref{SAcont}) conserves the total norm and Hamiltonian, which are
continuum counterparts of expressions (\ref{eq:Norm}) and (\ref{HSM}):
\begin{eqnarray}
\left( N_{\mathrm{AL}}\right) _{\mathrm{cont}} &=&\frac{1}{\mu }%
\int_{-\infty }^{+\infty }dx~{\ln }\left\vert 1+\mu |\Psi |^{2}\right\vert ,
\label{NSMcont} \\
\left( H_{\mathrm{SM}}\right) _{\mathrm{cont}} &=&\int_{-\infty }^{+\infty
}dx\left[ \left\vert \Psi _{x}\right\vert ^{2}-2\left( \frac{1}{\mu }%
+1\right) |\Psi |^{2}\right] +\frac{2}{\mu }\left( N_{\mathrm{AL}}\right) _{%
\mathrm{cont}}\,.  \label{SMcont}
\end{eqnarray}

It is relevant to mention that the general approximation opposite to the
continuum limit is the anti-continuum limit \cite{Aubry,Aubry2}. This
approach starts with the limit form of the DNLS equation, in which the
linear couplings between adjacent sites are dropped. Then, one can try to
construct various states, including solitons, by introducing an input
composed of simple solutions of the single-site equations corresponding to
Eq. (\ref{DNLSE}), \textit{viz}., $i\dot{\psi}_{l,m,n}=\sigma \left\vert
\psi _{l,m,n}\right\vert ^{2}\psi _{l,m,n}+V_{l,m,n}\psi _{l,m,n}$, at a
finite set of sites, and keeping the zero solution at all others. The
single-site \textquotedblleft simple solutions" are%
\begin{equation}
\psi _{l,m,n}=a_{l,m,n}\exp \left( -i\left\vert a_{l,m,n}\right\vert
^{2}t-iV_{l,m,n}t\right) ,  \label{anti}
\end{equation}%
where $a_{l,m,n}$ is an arbitrary set of complex amplitudes. Next, one
reintroduces weak linear intersite couplings and attempts to identify
nontrivial solutions that may thus appear from the finite-set input composed
of the single-site solutions (\ref{anti}).

\paragraph{Self-trapping in lattices with the self-repulsion strength
growing from the center to periphery}

DNLS equations with the onsite self-repulsive nonlinearity, corresponding to
$\sigma >0$ in Eq. (\ref{DNLSE}), may support discrete-soliton
(self-trapped) states without the resort to the staggering transform (\ref%
{stagg}) if the local self-repulsion strength is made a function of the
lattice coordinates, growing fast enough from the center to periphery.
Originally, this option was elaborated in the framework of the continuum NLS
and GP equations in the space of dimension $D$, with the local
self-defocusing (repulsion) coefficients growing at $r\rightarrow \infty $ ($%
r$ is the radial coordinate) faster than $r^{D}$ \cite{Borovkova}. In terms
of the 1D and 2D DNLS equations, similar settings were introduced in Refs.
\cite{Gligorich} and \cite{center-periphery}, with the site-dependent
self-attraction coefficients, $\left( \sigma _{n}\right) _{\mathrm{1D}%
}=\sigma _{0}\exp \left( \alpha |n|\right) $ and $\left( \sigma
_{m,n}\right) _{\mathrm{2D}}=\sigma _{0}\exp \left( \alpha \left(
|m|+|n|\right) \right) $, respectively, with positive constants $\sigma _{0}$
and $\alpha $. In \ the 2D model, solutions were constructed, and their
stability analyzed, for fundamental, dipole, quadrupole, and vortical
discrete solitons \cite{Borovkova}.

\paragraph{DNLS equations with long-range dipole-dipole and
quadrupole-quadrupole intersite interactions}

It is well known that atomic BEC formed of ultracold atoms carrying
permanent magnetic moments feature specific dynamical effects due to the
long-range interactions between atomic moments \cite{DD-review}. This fact
suggests to combine the dipole-dipole interactions and a deep OL potential,
thus introducing DNLS equations with the nonlocal (long-range) coupling
between lattice sites. In the 2D setting, this model gives rise to different
forms of the DNLS equations. The simplest setup corresponds to the case when
the atomic moments are polarized by external dc magnetic field perpendicular
to the system's plane. In this case, the dipole-dipole interactions amount
to the isotropic nonlocal repulsion, accounted for by the respective
interaction coefficient $\Gamma >0$ \cite{DD-2D}

\begin{eqnarray}
i\frac{\partial \psi _{m,n}}{\partial t} &=&-\frac{1}{2}\left( \psi
_{m+1,n}+\psi _{m-1,n}+\psi _{m,n+1}+\psi _{m,n-1}-4\psi _{m,n}\right) +%
\left[ \sigma \left\vert \psi _{m,n}\right\vert ^{2}\right.  \notag \\
&&\left. +\Gamma \sum_{\left\{ m^{\prime },n^{\prime }\right\} \neq \left\{
m,n\right\} }\left\vert \left( m-m^{\prime }\right) ^{2}+\left( n-n^{\prime
}\right) ^{2}\right\vert ^{-3/2}\left\vert \psi _{m^{\prime },n^{\prime
}}\right\vert ^{2}\right] \psi _{m,n}~,  \label{DGPE}
\end{eqnarray}%
where $\sigma $ is the same coefficient of the onsite self-interaction as in
Eq. (\ref{DNLSE}). A more sophisticated setup corresponds to the atomic
magnetic moments polarized parallel to the system's plane. In the former
case, the nonlocal term term in the respective DNLS equation is anisotropic,
being attractive in one in-plane direction and repulsive in the other, cf.
Ref. \cite{Ami1}:%
\begin{gather}
i\frac{\partial \psi _{m,n}}{\partial t}=-\frac{1}{2}\left( \psi
_{m+1,n}+\psi _{m-1,n}+\psi _{m,n+1}+\psi _{m,n-1}-4\psi _{m,n}\right) +%
\left[ \sigma \left\vert \psi _{m,n}\right\vert ^{2}\right.  \notag \\
\left. +\Gamma \sum_{\left\{ m^{\prime },n^{\prime }\right\} \neq \left\{
m,n\right\} }\frac{\left( n-n^{\prime }\right) ^{2}-2\left( m-m^{\prime
}\right) ^{2}}{\left[ \left( m-m^{\prime }\right) ^{2}+\left( n-n^{\prime
}\right) ^{2}\right] ^{5/2}}\left\vert \psi _{m^{\prime },n^{\prime
}}\right\vert ^{2}\right] \psi _{m,n}~.  \label{DDaniso}
\end{gather}%
The analysis reported in Ref. \cite{DD-2D} demonstrates that the nonlocal
repulsion, added to Eq. (\ref{DGPE}), helps to stabilize discrete solitons
with embedded vorticity. Solutions of Eq. (\ref{DDaniso}) for anisotropic
vortex solitons can be found too, but they are completely unstable \cite%
{DD-2D}.

A 2D DNLS model which combines the local onsite nonlinearity and long-range
interaction between particles carrying permanent quadrupole electric moments
was elaborated in Ref. \cite{QQ}:%
\begin{gather}
i\frac{\partial \psi _{m,n}}{\partial t}=-\frac{1}{2}\left( \psi
_{m+1,n}+\psi _{m-1,n}+\psi _{m,n+1}+\psi _{m,n-1}-4\psi _{mn}\right) +\left[
\sigma |\psi _{mn}|^{2}\right.  \notag \\
\left. +\Gamma \sum_{\left\{ m^{\prime },n^{\prime }\right\} \neq \left\{
m,n\right\} }\frac{(n-n^{\prime })^{2}-4(m-m^{\prime })^{2}}{[(m-m^{\prime
})^{2}+(n-n^{\prime })^{2}]^{7/2}}|\psi _{m^{\prime }n^{\prime }}|^{2}\right]
\psi _{m,n},  \label{QQ}
\end{gather}%
cf. Eq. (\ref{DGPE}). This model also gives rise to families of stable 2D
discrete solitons \cite{QQ}.

\paragraph{The 2D discrete second-harmonic-generating ($\protect\chi ^{(2)}$%
) system}

The quadratic (alias $\chi ^{(2)}$) nonlinearity is a fundamentally
important effect which gives rise to coherent generation of the second
harmonic in optics. In terms of the 2D spatial-domain propagation in a
continuum material, the standard $\chi ^{(2)}$ system for amplitudes $\psi
\left( x,y,z\right) $ and $\phi \left( x,y,z\right) $ of the
fundamental-frequency (FF) and second-harmonic (SH) waves is \cite{Buryak}%
\begin{eqnarray}
i\psi _{z} &=&-(1/2)\nabla ^{2}\psi -\psi ^{\ast }\phi \,,  \notag \\
2i\phi _{z} &=&-(1/2)\nabla ^{2}\phi -Q\phi -\frac{1}{2}\psi ^{2}\,,
\label{SHG}
\end{eqnarray}%
where $z$ is the propagation distance, the paraxial-diffraction operator $%
(1/2)\nabla ^{2}$ acts on the transverse coordinates $\left( x,y\right) $, $%
Q $ is a real mismatch parameter, and $\ast $ stands for the complex
conjugate. The discretized version of Eqs. (\ref{SHG}), which represents, in
the tightly-binding approximation, light propagation in a photonic crystal
made of the $\chi ^{(2)}$ material, is \cite{Hadi}

\begin{eqnarray}
i\frac{d\psi _{m,n}}{dz} &=&-C_{1}\left( \psi _{m+1,n}+\psi _{m-1,n}+\psi
_{m,n+1}+\psi _{m,n-1}-4\psi _{mn}\right) -\psi _{m,n}^{\ast }\phi _{m,n}\,,
\notag \\
2i\frac{d\phi _{m,n}}{dz} &=&-\frac{C_{2}}{2}\left( \phi _{m+1,n}+\phi
_{m-1,n}+\phi _{m,n+1}+\phi _{m,n-1}-4\phi _{mn}\right) -Q\phi _{m,n}-\frac{1%
}{2}\psi _{m,n}^{2}\,,  \label{geq4}
\end{eqnarray}%
where $C_{1}$ and $C_{2}$ are effective lattice-coupling constants for the
FF and SH waves. The role of the conserved norm of the discrete $\chi ^{(2)}$
system is played by the Manley-Rowe invariant, i.e., the total optical
power, $I=\sum_{m,n}\left( \left\vert \psi _{m,n}\right\vert
^{2}+4\left\vert \phi _{m,n}\right\vert ^{2}\right) $.

An essential property of 2D discrete solitons produced by Eqs. (\ref{geq4})
is their mobility \cite{Hadi}. In this connection, it is relevant to mention
that, while the development of the quasi-collapse in the 2D discrete NLS
equation with the cubic self-attraction is arrested by the underlying
lattice structure, the quasi-collapse strongly pins the 2D solitons to the
same structure, and thus makes them immobile. On the other hand, the $\chi
^{(2)}$ nonlinearity does not give rise to the collapse in the 2D (and 3D)
space, therefore 2D $\chi ^{(2)}$ solitons do not demonstrate a trend for
strong pinning, remaining effectively mobile robust localized modes \cite%
{Hadi}.

\subsection{One-dimensional DNLS solitons}

\subsubsection{Fundamental solitons}

In the 1D setting, the model of basic interest is the DNLS equation with
self-attraction, which corresponds to $\sigma =-1$ in the 1D version of Eq. (%
\ref{DNLSE}), without the external potential ($V_{l,m,n}=0$):%
\begin{equation}
i\dot{\psi}_{n}=-(1/2)\left( \psi _{n+1}+\psi _{n-1}-2\psi _{n}\right)
-\left\vert \psi _{n}\right\vert ^{2}\psi _{n}\ .  \label{1D-DNLSE}
\end{equation}%
This equation conserves two dynamical invariants, \textit{viz}., the total
norm,%
\begin{equation}
N=\sum_{n=-\infty }^{+\infty }\left\vert \psi _{n}\right\vert ^{2},
\label{Ndiscr}
\end{equation}%
and Hamiltonian (energy),%
\begin{equation}
H=\sum_{n=-\infty }^{+\infty }\left[ (1/2)\left\vert \psi _{n}-\psi
_{n-1}\right\vert ^{2}-(1/4)\left\vert \psi _{n}\right\vert ^{4}\right] .
\label{H-DNLSE}
\end{equation}

Stationary solutions to Eq. (\ref{1D-DNLSE}) with real frequency $\omega $
are looked for as
\begin{equation}
\psi _{n}(t)=e^{-i\omega t}u_{n},  \label{psi-u}
\end{equation}%
with real amplitudes $u_{n}$ satisfying the discrete equation,
\begin{equation}
\omega u_{n}=-(1/2)\left( u_{n+1}+u_{n-1}-2u_{n}\right) -u_{n}^{3}.
\label{1D-DNLS-stat}
\end{equation}%
Note that Eq. (\ref{1D-DNLS-stat}) can be derived by varying its Lagrangian,%
\begin{equation}
L=\sum_{n=-\infty }^{+\infty }\left\{ \frac{1}{4}\left[ \left(
u_{n}-u_{n-1}\right) ^{2}-u_{n}^{4}\right] -\frac{\omega }{2}%
u_{n}^{2}\right\} ,
\label{LagrDNLS}
\end{equation}%
with respect to the discrete real field $u_{n}$.

\begin{figure}
\includegraphics[width=7.5cm]{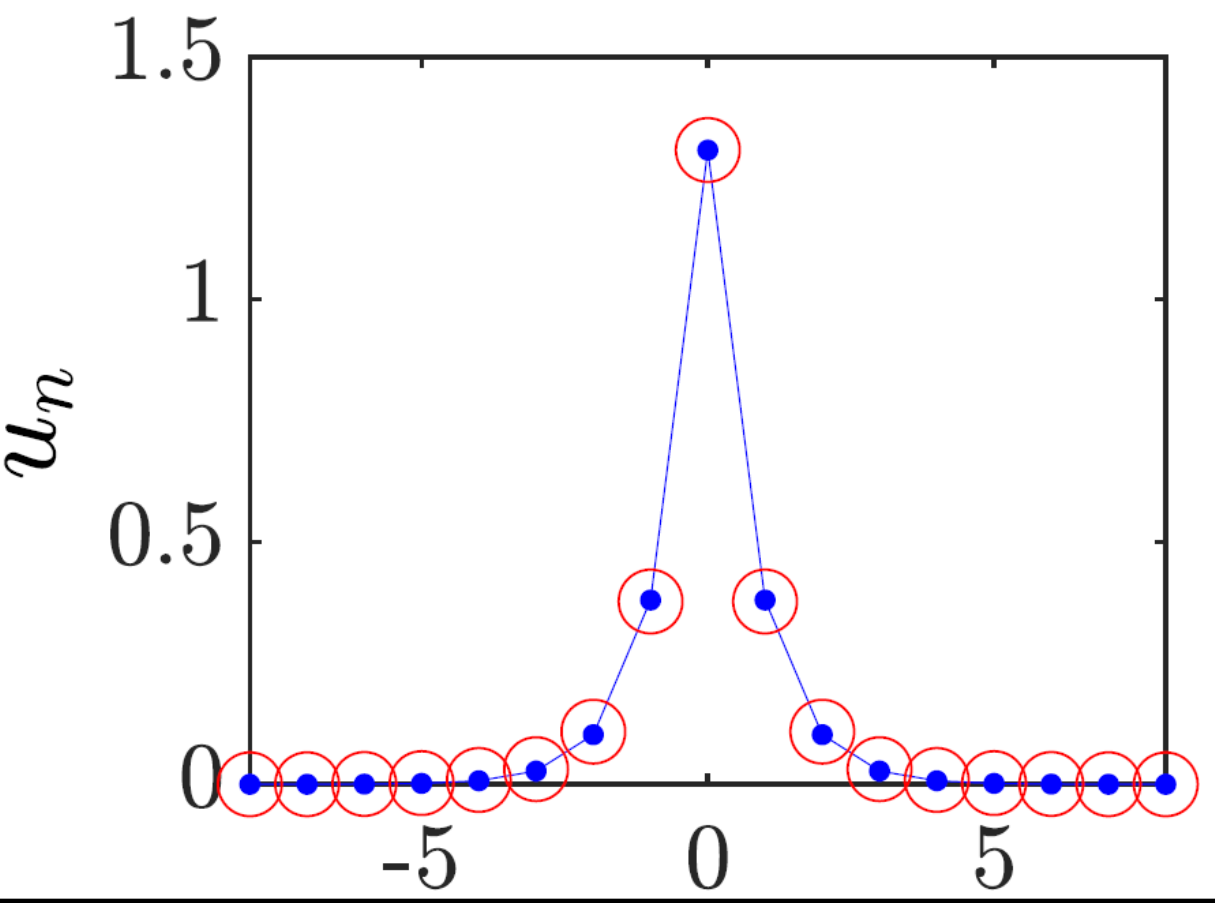}
\caption{Comparison of a typical discrete soliton, obtained as the numerical
solution of Eq. (\protect\ref{1D-DNLS-stat}), shown by chains of blue dots,
and its counterpart produced by the VA (shown by red open circles). In
this figure, $\protect\omega =-1$, see Eq. (\protect\ref{psi-u}), the
corresponding parameters of ansatz (\protect\ref{ansatz-onsite}) being $%
A\approx 1.31$, $a\approx 1.15$. The figure is
borrowed from Ref. \protect\cite{chapter}.}
\label{fig1}
\end{figure}

A fundamental property of the DNLS equation (\ref{1D-DNLSE}) with the
self-attractive onsite nonlinearity is the modulational instability (MI) of
its spatially homogeneous continuous-wave (CW) state \cite{Peyrard}, $\psi
_{n}=a\exp \left( ia^{2}t\right) $, with an arbitrary amplitude $a$ [cf. Eq.
(\ref{psi-u})]. MI breaks the CW state into a chain of discrete solitons
\cite{discr-review}. Analytical solutions for these solitons are not
available, as the DNLS equation is not integrable. The solitons can be
readily found in a numerical form, and studied in the framework of the
variational approximation (VA) \cite{DNLS-book}. The VA is based on a
particular \textit{ansatz}, i.e. an analytical expression which aims to
approximate the solution \cite{Progress}. The only discrete ansatz for which
analytical calculations are feasible is represented by the exponential
function \cite{Weinstein,Papa,Dave-VA,Gorder}, namely,%
\begin{equation}
\left( u_{n}\right) _{\mathrm{onsite}}=A\exp \left( -a|n|\right) ,
\label{ansatz-onsite}
\end{equation}%
with $a>0$. The corresponding norm, calculated as per Eq. (\ref{Ndiscr}), is%
\begin{equation*}
N_{\mathrm{ansatz}}=A^{2}\coth a\text{.}
\end{equation*}%
Note that ansatz (\ref{ansatz-onsite}) is appropriate for strongly and
moderately discrete solitons, as shown in Fig. \ref{fig1}, but not for broad
(quasi-continuum) ones, which may be approximated by the commonly known
soliton solution of the NLS equation (the 1D version of (\ref{NLSE}) with $%
U=0$),%
\begin{equation}
\psi \left( x,t\right) =\eta ~\mathrm{sech}\left( \eta \left( x-\xi \right)
\right) \exp \left( i\eta ^{2}t\right) ,  \label{cont-soliton}
\end{equation}%
with width $\eta ^{-1}$ which must be large in comparison with the
discreteness spacing, $\eta ^{-1}\gg 1$, and central coordinate $\xi $.

The substitution of ansatz (\ref{ansatz-onsite}) in Lagrangian (\ref%
{LagrDNLS}) produces the corresponding VA Lagrangian:%
\begin{equation}
L_{\mathrm{VA}}=(A^{2}/2)\tanh (a/2)-(A^{4}/4)\coth \left( 2a\right)
-(\omega /2)A^{2}\coth a.  \label{Leff}
\end{equation}%
Then, for given $\omega <0$ (solitons do not exist for $\omega >0$), the
squared amplitude, $A^{2}$, and inverse width, $a$, of the discrete soliton
are predicted by the Euler-Lagrange equations,%
\begin{equation}
\frac{\partial L_{\mathrm{eff}}}{\partial \left( A^{2}\right) }=\frac{%
\partial L_{\mathrm{eff}}}{\partial a}=0.  \label{EulerLagr}
\end{equation}%
This corresponding system of algebraic equations for $A^{2}$ and $a$ can be
easily solved numerically. The VA produces an accurate predictions for the
solitons, as shown in Fig.~\ref{fig1} and Ref.~\cite{Needs07}.

Rigorous justification of the VA was elaborated in Ref. \cite{rigorous}.
Furthermore, the VA and a full numerical solution of Eq. (\ref{1D-DNLSE})
demonstrate that the entire family of the discrete solitons is stable \cite%
{DNLS-book}.

In addition to the bright solitons considered here, the DNLS equation also
gives rise to discrete dark solitons, which have been studied in detail
theoretically and experimentally \cite{Kivshar1,Konotop,Silberberg2,Pengfei}.
As concerns the topic of the present review, two-dimensional discrete dark
modes, such as delocalized lattice vortices, were studied too \cite%
{Kevrekidis}. However, the consideration of dark modes is not included in
this article.

\subsubsection{Higher-order one-dimensional modes: twisted discrete solitons
and bound states}

In addition to the fundamental (single-peak) solitons outlined above, Eq. (%
\ref{1D-DNLS-stat}) admits stable second-order states in the form of \textit{%
twisted modes}, which are subject to the antisymmetry condition, $%
u_{n}=-u_{1-n}$ \cite{twisted}. Such states exist and are stable only in a
strongly discrete form, vanishing in the continuum limit. In particular, the
above-mentioned anti-continuum approximation is appropriate for the
construction of the twisted modes.

Stable 1D DNLS solitons may form bound states, which also represent
higher-order modes of the DNLS equation. They are stable in the \textit{%
out-of-phase} form, i.e., with opposite signs of the constituent solitons
\cite{bound states,bound states 2}, which resembles the structure of the
twisted modes (however, the tight antisymmetric structure of the twisted
modes cannot be considered as a bound state of fundamental solitons), the
same being true for 2D DNLS solitons \cite{Bishop}. Note that stationary
bound states of fundamental solitons do not exist in the continuum limit.

\subsubsection{1D solitons in the Salerno model (SM)}

The substitution of the usual ansatz (\ref{psi-u}) in Eq. (\ref{SAmodel})
produces a stationary discrete equation for real $u_{n}$:%
\begin{equation}
\omega u_{n}=-\left( u_{n+1}+u_{n-1}\right) \left( 1+\mu u_{n}^{2}\right)
-2u_{n}^{3},  \label{1D-SM-stat}
\end{equation}%
cf. (\ref{1D-DNLS-stat}). Discrete solitons produced by the SM equation (\ref%
{SAmodel}) with $\mu >0$, i.e. with \emph{noncompeting} intersite and onsite
self-focusing nonlinearities, were investigated by means of numerical
methods \cite{Cai, Cai97, Dmitriev03}. It was found that the SM\ gives rise
to the 1D solitons at all positive values of $\mu $.

Another option is to consider the SM with $\mu <0$, which features \emph{%
competing nonlinearities}, as the intersite cubic term, with coefficient $%
\mu <0$ in Eq. (\ref{SAmodel}), which accounts for the nonlinear coupling
between adjacent sites of the lattice, and the onsite cubic term in Eq. (\ref%
{SAmodel}) represent, respectively, repulsive and attractive nonlinear
interactions. This version of the SM gives rise to families of discrete
solitons, in the usual form (\ref{psi-u}), with $\omega <0$ and real $u_{n}$%
, of two different types.\ One family represents ordinary discrete solitons,
similar to those generated by the DNLS equation. Another family represents
\textit{cuspons, }featuring higher curvature of their profile at the center.
A small subfamily of ordinary solitons produced by the SM with the competing
nonlinearities is unstable, while all cuspons are stable.

As mentioned above, antisymmetric bound states of DNLS solitons are stable,
while symmetric bound states are unstable \cite{bound states,bound states 2}%
. The same is true for bound states of ordinary discrete solitons in the SM
\cite{Zaragoza}. However, in the framework of the SM with the competing
nonlinearities, the situation is \emph{exactly opposite} for the cuspons:
their symmetric and antisymmetric bound states are stable and unstable,
respectively \cite{Zaragoza}.

\subsection{The subject and structure of the present article}

The above-mentioned reviews \cite{Aubry,discr-review,Rothos,Tsoy,chapter}
and \cite{DNLS-book} produce a comprehensive survey of theoretical and
experimental results for discrete solitons in various 1D systems. The
objective of this article is to produce a relatively brief summary of
results for multidimensional (chiefly, two-dimensional) discrete and
semi-discrete solitons, which were considered in less detail in previous
reviews and, on the other hand, draw growing interest in the context of the
current work with 2D and 3D solitons in diverse physical contexts \cite%
{NRP,book}. In this context, the presence of the two or three coordinates
makes it also possible to define \textit{semi-discrete} states as ones which
are discrete in one direction and continuous in the perpendicular one \cite%
{Blit,Driben,Raymond,semidiscr vort Adv Phot,Raymond2}. The article chiefly
represents theoretical results, but some experimental findings for
quasi-discrete 2D solitons in photonic lattices \cite{Kivshar,Segev} are
included too.

The review presented below does not claim to be comprehensive. It comprises
results that are produced by conservative models of the DNLS types
(including, in particular, the 2D SM). Discrete models of other types -- in
particular, those similar to the Toda lattice, see Eq. (\ref{TLEu}),
Fermi-Pasta-Ulam-Tsingou lattices \cite{FPUT}, and Frenkel-Kontorova systems
\cite{FK} -- are not considered here. Dissipative systems are not considered
either, except for a 2D model with the parity-time ($\mathcal{PT}$) symmetry
\cite{PT}, see Eq. (\ref{basicEq}) below.

The rest of the article is arranged as follows. Basic results for
fundamental (zero-vorticity) and vortex solitons, as well as bound states of
such solitons, produced by the 2D DNLS equation and its generalizations, are
summarized in Section II, which is followed, in Section III, by brief
consideration of fundamental and vortex solitons in the 2D SM (Salerno
model). Section IV addresses discrete solitons of the semi-vortex and
mixed-mode types in the 2D spin-orbit-coupled (SOC) system of GP equations
for a two-component BEC. Basic results for discrete self-trapped modes
produced by 3D DNLS equations, including fundamental and vortex solitons,
along with skyrmions, are presented in Section V. The findings for 2D
\textit{semi-discrete} systems, again including fundamental and vortex
solitons, supported by combined quadratic-cubic and cubic-quintic
nonlinearities (that are relevant for BEC and optics, respectively), are
summarized in Section VI. This section also addresses transverse mobility of
confined spatiotemporal modes in an array of optical fibers with the
intrinsic cubic self-focusing (Kerr nonlinearity). Fundamental and vortex
solitons produced by a $\mathcal{PT}$-symmetric discrete 2D coupler with the
cubic nonlinearity are considered in Section VII. The article is concluded
by Section VIII, which, in particular, suggests directions for the further
work in this area, and mentions particular topics which are not included in
the present review.

\section{Two-dimensional (2D) nonlinear-Schr\"{o}dinger lattices:
fundamental and vortex solitons, and their bound states}

\subsection{Vortex solitons: theoretical and experimental results}

The basic 2D cubic DNLS equation is the 2D version of Eq. (\ref{DNLSE}) with
the self-attraction ($\sigma =-1$) and without the external potential:%
\begin{equation}
i\dot{\psi}_{m,n}=-(1/2)\left( \psi _{m+1,n}+\psi _{m-1,n}+\psi
_{m,n+1}+\psi _{m,n-1}-4\psi _{m,n}\right) -\left\vert \psi
_{m,n}\right\vert ^{2}\psi _{m,n},  \label{2D}
\end{equation}%
cf. Eq. (\ref{1D-DNLSE}). The substitution of $\psi _{m,n}=\exp \left(
-i\omega t\right) u_{m,n}$ in Eq. (\ref{2D}) produces the stationary
equation, where the stationary discrete wave function, $u_{m,n}$, may be
complex:
\begin{equation}
\omega u_{m,n}=-(1/2)\left(
u_{m+1,n}+u_{m-1,n}+u_{m,n+1}+u_{m,n-1}-4u_{m,n}\right) -\left\vert
u_{m,n}\right\vert ^{2}u_{m,n},  \label{2D-DNLSE}
\end{equation}%
cf. (\ref{1D-DNLS-stat}). Fundamental-soliton solutions to Eq. (\ref%
{2D-DNLSE}) can be predicted by means of VA \cite{Weinstein-2D,Chong}, using
an exponential ansatz, see (\ref{SSB-ansatz}) below (cf. Eq. (\ref%
{ansatz-onsite}) for the 1D soliton). More interesting in the 2D setting are
discrete solitons with \textit{embedded vorticity}, which were introduced in
Ref. \cite{we} (see also Ref. \cite{they}). Vorticity, alias the topological
charge, or winding number, is defined as $\Delta \varphi /\left( 2\pi
\right) $, where $\Delta \varphi $ is a total change of the phase of the
complex discrete function $u_{m,n}$ along a contour surrounding the vortex'
pivot. Stability is an important issue for 2D discrete solitons, because, in
the continuum limit, the 2D\ NLS equation gives rise to the well-known
Townes solitons \cite{Townes}, which are unstable against the onset of the
critical collapse \cite{Gadi}. In the same limit, the Townes solitons with
embedded vorticity (vortex rings \cite{Minsk}) are subject to much stronger
instability against spontaneous splitting of the ring in fragments \cite%
{PhysD}.

\begin{figure}
\includegraphics[width=10cm,scale=1]{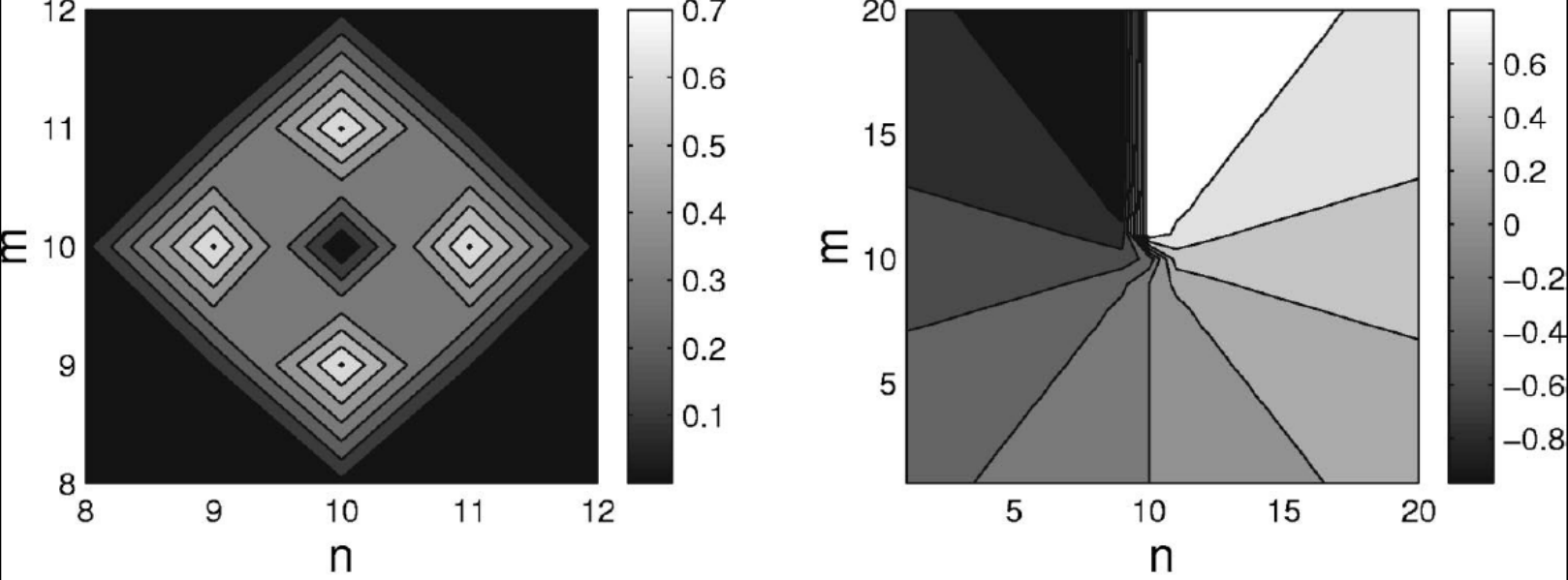}
\caption{A stable discrete vortex soliton with topological charge $S=1$,
produced by Eq. (\protect\ref{2D-DNLSE}) with $\protect\omega =-3.2$. The
left and right panels show, respectively, distributions of the absolute
value and phase of the complex wave function $u_{m,n}$ in the plane with
coordinates $\left( m,n\right) $. The figure is borrowed from Ref.
\protect\cite{we}.}
\label{fig2}
\end{figure}

The lattice structure of the DNLS equation provides for stabilization of
both fundamental (zero-vorticity) and vortex solitons \cite{we}. A typical
example of a stable 2D vortex soliton with topological charge $S=1$ is
displayed in Fig. \ref{fig2}. 2D fundamental and vortex solitons, with
topological charges $S=0$ and $1$, are stable in regions $-\omega >|\omega _{%
\mathrm{cr}}^{(S=0)}|\approx 0.50$ \ and $-\omega >|\omega _{\mathrm{cr}%
}^{(S=1)}|\approx 1.23$, respectively \cite{we}, while the higher-order
discrete vortices with charges $S=2$ and $4$ are unstable, being replaced by
stable modes in the form of quadrupoles and octupoles \cite{Zhigang}. The
vortex solitons with $S=3$ may be stable, but only in an extremely discrete
form, \textit{viz}., at $-\omega >|\omega _{\mathrm{cr}}^{(S=2)}|\approx
4.94 $. In agreement with what is said above, these results imply that all
the solitons are unstable in the continuum limit, corresponding, in the
present notation, to $\omega \rightarrow 0$.

The experimentally relevant lattice structure may be anisotropic, with the
linear combination $\left( \psi _{m+1,n}+\psi _{m-1,n}+\psi _{m,n+1}+\psi
_{m,n-1}-4\psi _{m,n}\right) $ in Eq. (\ref{2D}) replaced by

\noindent $\left( \alpha \left( \psi _{m+1,n}+\psi _{m-1,n}\right) +\left(
\psi _{m,n+1}+\psi _{m,n-1}\right) -2\left( 1+\alpha \right) \psi
_{m,n}\right) ,$ with anisotropy parameter $\alpha \neq 1$. Effects of the
anisotropy on the structure and stability of the fundamental and vortical
discrete solitons were explored in Ref. \cite{aniso}.

The theoretically predicted 2D discrete solitons with vorticity $S=1$ were
experimentally created in Refs. \cite{Kivshar} and \cite{Segev}, using a
photorefractive crystal. Unlike uniform media of this type, where
delocalized (\textquotedblleft dark") optical vortices were originally
demonstrated \cite{Zhig1,Zhig2}, those works made use of a deep virtual
photonic lattice as a quasi-discrete structure supporting the self-trapping
of nonlinear modes in the optical field with extraordinary polarization
(while the photonic lattice was induced as the interference pattern of
quasi-linear beams in the ordinary polarization). Intensity distributions
observed in vortex solitons of the onsite- and intersite-centered types
(i.e., with the vortex' pivot coinciding with a site of the underlying
lattice, or set between the sites, respectively), are displayed in Fig. \ref%
{fig3}.

\begin{figure}
\includegraphics[width=10cm,scale=1]{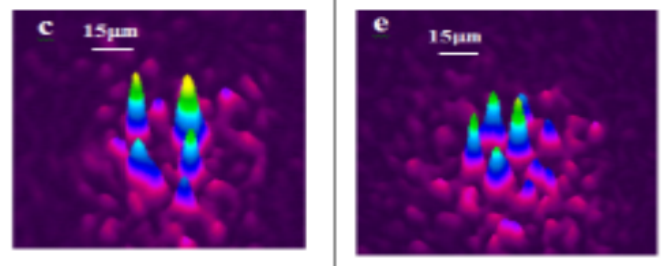}
\caption{Quasi-discrete optical solitons with vorticity $S=1$, created in
the photorefractive crystal with an induced deep photonic lattice. The left
and right panels display, respectively, the onsite- and intersite-centered
vortex solitons. The figure is borrowed from Ref. \protect\cite{Segev}.}
\label{fig3}
\end{figure}
Another interesting result demonstrated (and theoretically explained) in
deep virtual photonic lattices is a possibility of periodic flipping of the
topological charge of a vortex soliton initially created with topological
charge $S=2$ \cite{Chen}. Stable vortex solitons with $S=2$ were created
using a hexagonal virtual photonic lattice (while, as mentioned above, the
localized modes with $S=2$ are completely unstable in the case of the
square lattice) \cite{Denz}.

\subsection{Bound states of 2D discrete solitons and solitary vortices}

\begin{figure}
\includegraphics[width=12cm,scale=1]{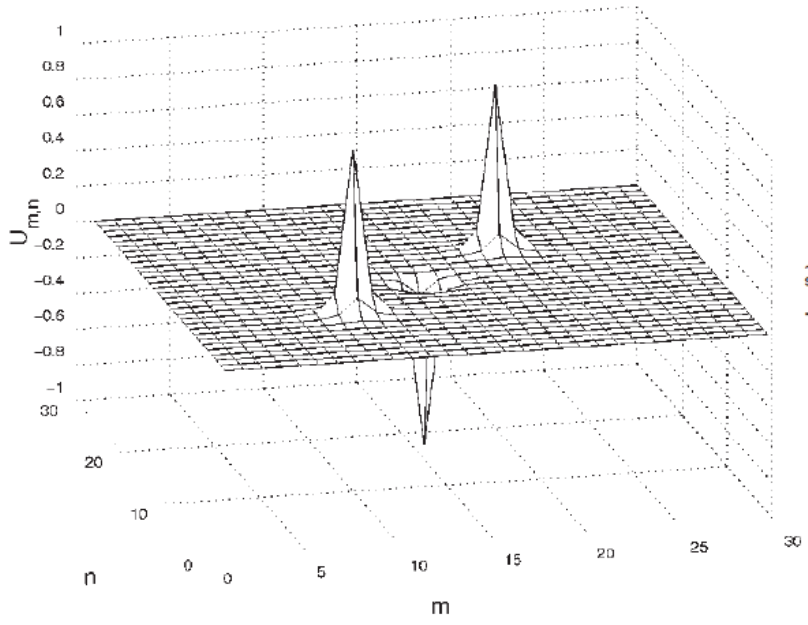}
\caption{An example of a stable bound state of three 2D fundamental
(zero-vorticity) solitons, which are mutually out of phase. The solution is
produced by the numerical solution of Eq. (\protect\ref{2D-DNLSE}) with $%
\protect\omega =-1$. The figure is borrowed from Ref. \protect\cite{Bishop}.
[This low-quality figure will be replaced in the published version of the paper.]}
\label{fig7}
\end{figure}

As mentioned above, stable 2D discrete solitons may form stable bound
states, composed of two or several items. Vortex solitons may form bound
states as well. This possibility and stability of the resulting bound states
are determined by an effective potential of interaction between two
identcial discrete solitons with intrinsic vorticity $S$, which are
separated by large distance $L$. The potential can be derived from an
asymptotic expression for exponentially decaying tails of the soliton. In
the quasi-continuum approximation, it is $u_{m,n}\sim \left(
m^{2}+n^{2}\right) ^{-1/4}\exp \left( -\sqrt{-2\omega \left(
m^{2}+n^{2}\right) }\right) $ (recall soliton solutions to Eq. (\ref%
{2D-DNLSE}) exist for $\omega <0$). Then, the overlap of the tail of each
soliton with the central body of the other one gives rise to the following
interaction potential:%
\begin{equation}
U_{\mathrm{int}}(L)\approx \mathrm{const}\cdot \cos \delta \cdot
(-1)^{S}L^{-1/2}\exp \left( \sqrt{-2\omega }L\right) ,  \label{Uint}
\end{equation}%
with $\mathrm{const}>0$, where $\delta $ is the phase shift between the
solitons. Thus, for the fundamental solitons with $S=0$, Eq. (\ref{Uint})
predicts the attractive interaction between the in-phase solitons ($\delta =0
$), and repulsion between out-of-phase ones ($\delta =\pi $). Accordingly,
the interplay of the repulsive interaction with the effective
Peierls-Nabarro potential, which is pinning the soliton to the underlying
lattice \cite{DNLS-book}, produces \emph{stable} bound states of two or
several mutually out-of-phase solitons, while the in-phase bound states are
unstable. These predictions were confirmed by numerical results \cite{Bishop}%
. As an example, Fig. \ref{fig7} displays a numerically found stable bound
state in the form of a string of three fundamental solitons with alternating
phases.

\begin{figure}
\centering
\includegraphics[width=15cm,scale=1]{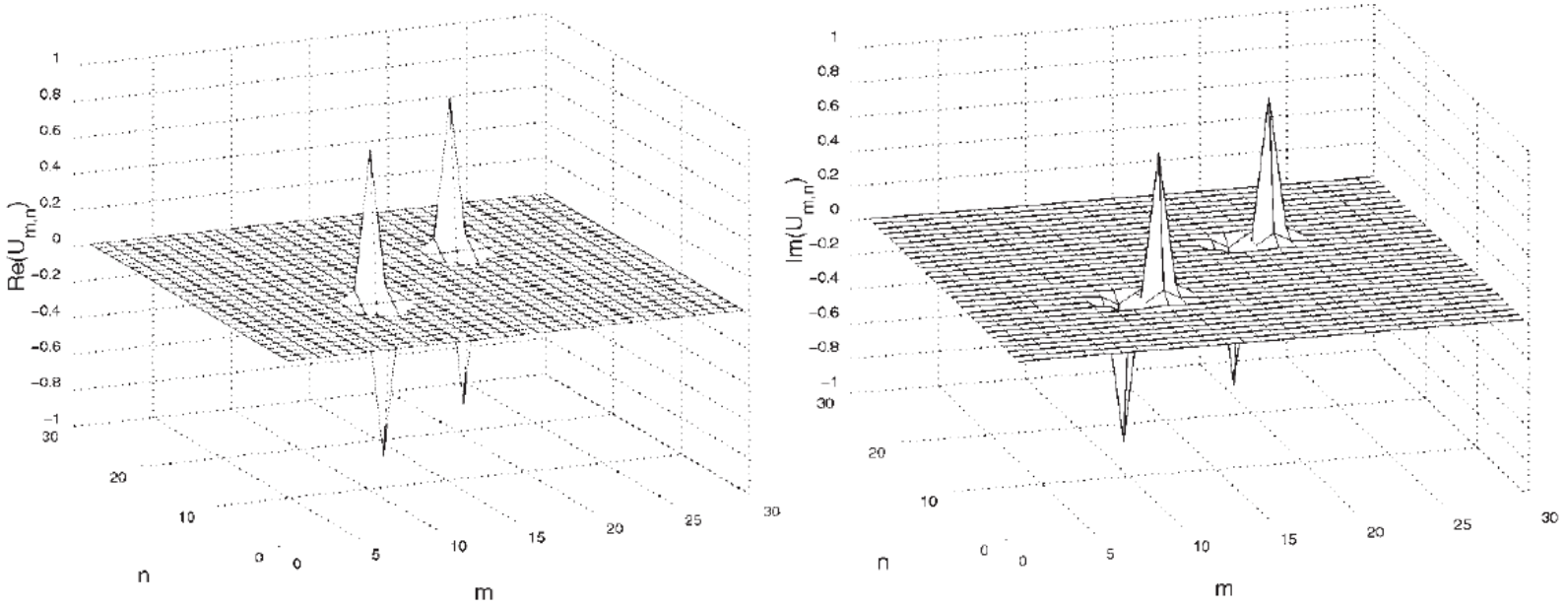}
\caption{An example of a stable bound state of two in-phase discrete vortex
solitons, with topological charge $S=1$. The result is produced by the
numerical solution of Eq. (\protect\ref{2D-DNLSE}) with $\protect\omega =-1$%
. The figure is borrowed from Ref. \protect\cite{Bishop}
[This low-quality figure will be replaced in the published version of the paper.].}
\label{fig8}
\end{figure}

For the pair of identical vortex solitons with $S=1$, Eq. (\ref{Uint})
predicts the opposite result, \textit{viz}., the repulsive interaction and
stability of the ensuing bound states for in-phase vortices ($\delta =0$),
and the attraction leading to instability of the bound state in the case of $%
\delta =\pi $. These predictions were also corroborated by numerical
findings \cite{Bishop}, see an example of a stable bound state of two
identical vortex solitons in Fig. \ref{fig8}.

\subsection{2D discrete solitons in mini-gaps of a spatially modulated
lattice}

A specific class of self-trapped modes are gap solitons which may populate
finite bandgaps in linear spectra of various nonlinear systems originating
in optics and BEC \cite{Sterke,Brazhnyi,Morsch}. While in most cases gap
solitons are predicted theoretically and created experimentally in the
context of matter waves \cite{Oberthaler} and optical pulses \cite{Mok} in
the continuum, they may naturally appear\ as discrete modes in \textit{%
mini-gaps}, which are induced in the linear spectrum of lattice media by
superimposed periodic spatial modulations (\textit{superlattice}).

Such a 2D lattice model was introduced in Ref. \cite{Belgrade}, based on the
following DNLS equation:%
\begin{equation}
i\dot{\psi}_{m,n}+C_{m,n}\left( \psi _{m+1,n}+\psi _{m-1,n}\right)
+K_{m,n}\left( \psi _{m,n+1}+\psi _{m,n-1}\right) +\left\vert \psi
_{m,n}\right\vert ^{2}\psi _{m,n}=0,  \label{eq1}
\end{equation}%
where the horizontal and vertical coupling constants are modulated as
follows:
\begin{equation}
C_{m,n}=1+\Delta \cos (Qm),~K_{m,n}=1+\Delta \cos (Qn),  \label{CK}
\end{equation}%
cf. Eq. (\ref{DNLSE}). The superlattice represented by Eqs. (\ref{eq1}) and (%
\ref{CK}) can be created by means of the technique used for making OLs in
experiments with BEC \cite{Belgrade}.

Looking for solutions in the usual form, $\psi _{m,n}(t)=\exp \left(
-i\omega t\right) u_{m,n}$, with real frequency (chemical potential) $\omega
$, the numerical analysis produces the linear spectrum of this system,
including the usual semi-infinite bandgap and a pair of additional narrow
mini-gaps. Further, a family of fundamental 2D discrete solitons populating
the mini-gaps was furnished by the numerical solution of the full nonlinear
system, being stable in a small section of the mini-gap, as shown in Fig. %
\ref{fig8}. The stable 2D soliton displayed in panel \ref{fig8}(a) features
a typical shape of gaps solitons, with a number of small satellite peaks
surrounding the tall central one \cite{Brazhnyi}. Bound states of two and
four fundamental solitons were found too, featuring weak instability \cite%
{Belgrade}.
\begin{figure}
\includegraphics[width=11cm,scale=1]{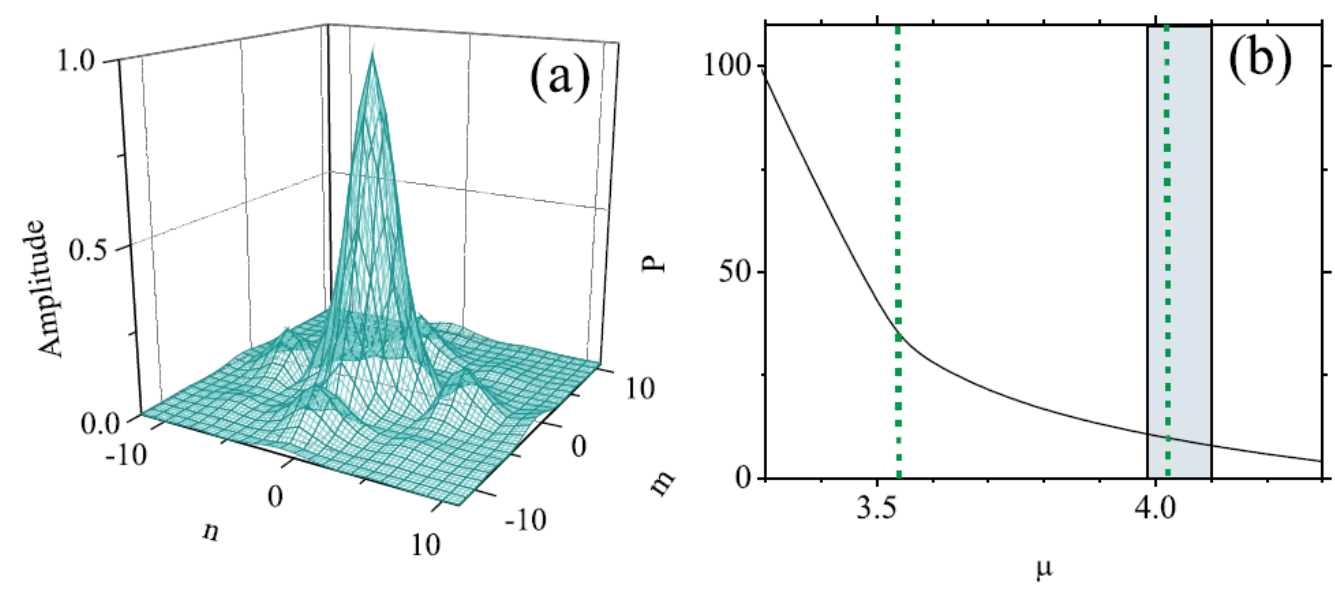}
\caption{(a) An example of a stable 2D discrete soliton with chemical
potential $\protect\mu =4.05$ (in this figure, the notation for the chemical potential is
$\mu$, instead of $\omega$, adopted in the text), which corresponds to the right vertical line
in (b), found in the mini-gap of the system based on Eqs. (\protect\ref{eq1}%
) and (\protect\ref{CK}) with $\Delta =0.5$ and $Q=\protect\pi /3$. (b) The
dependence $P(\protect\mu )$ of the norm of the solitons populating the
mini-gap, which is identical to the interval of values of $\protect\mu $
represented in the panel. The solitons are stable in the narrow shaded
interval. The figure is borrowed from Ref. \protect\cite{Belgrade}.}
\label{fig9}
\end{figure}

\subsection{2D discrete solitons in a rotating lattice}

Dynamics of BEC loaded in OLs rotating at angular velocity $\Omega $, as
well as the propagation of light in a twisted nonlinear photonic crystal
with pitch $\Omega $, is modeled by the 2D version of Eq. (\ref{NLSE})
including the lattice potential, with depth $\varepsilon $ and period $2\pi
/k$, written in the rotating reference frame:
\begin{equation}
i\psi _{t}=-\left[ (1/2)\nabla ^{2}+\Omega \hat{L}_{z}\right] \psi
-\varepsilon \left[ \cos \left( kx\right) +\cos (ky)\right] \psi +\sigma
|\psi |^{2}\psi ,  \label{rotating}
\end{equation}%
where $\hat{L}_{z}=i(x\partial _{y}-y\partial _{x})\equiv i\partial _{\theta
}$ is the operator of the $z$-component of the orbital momentum, $\theta $
being the angular coordinate in the $\left( x,y\right) $ plane. In the
tight-binding approximation, Eq. (\ref{rotating}) is replaced by the
following variant of the DNLS equation \cite{JC2}:
\begin{gather}
i\dot{\psi}_{m,n}=-(C/2)\left\{ \left( \psi _{m+1,n}+\psi _{m-1,n}+\psi
_{m,n+1}+\psi _{m,n-1}-4\psi _{m,n}\right) \right.  \notag \\
\left. -i\Omega \left[ m\left( \psi _{m,n+1}-\psi _{m,n-1}\right) -n\left(
\psi _{m+1,n}-\psi _{m-1,n}\right) \right] \right\} +\sigma |\psi
_{m,n}|^{2}\psi _{m,n}~,  \label{discrete}
\end{gather}%
where $C$ is the intersite coupling constant. In Ref. \cite{JC2}, stationary
solutions to Eq. (\ref{discrete}) were looked for in the usual form (\ref%
{psi-u}), fixing $\omega \equiv -1$ and varying $C$ in (\ref{discrete}) as a
control parameter. Two species of localized states were thus constructed:
off-axis fundamental discrete solitons, placed at distance $R$ from the
origin, and on-axis ($R=0$) vortex solitons, with topological numbers $S=1$
and $2$. At a fixed value of rotation frequency $\Omega $, a stability
interval for the fundamental soliton, $0<C<C_{\max }^{\mathrm{(fund)}}(R)$,
monotonously shrinks with the increase of $R$, i.e., most stable are the
discrete fundamental solitons with the center placed at the rotation pivot.
Vortices with $S=1$ are gradually destabilized with the increase of $\Omega $
(i.e., their stability interval, $0<C<C_{\max }^{\mathrm{(vort)}}(\Omega )$,
shrinks). On the contrary, a remarkable finding is that vortex solitons with
$S=2$, which, as said above, are completely unstable in the usual DNLS
equation with $\Omega =0$, are \emph{stabilized} by the rotation, in an
interval $0<C<C_{\mathrm{cr}}^{(S=2)}(\Omega )$, with $C_{\mathrm{cr}%
}^{(S=2)}(\Omega )$ growing as a function of $\Omega $. In particular, $C_{%
\mathrm{cr}}^{(S=2)}(\Omega )\approx 2.5\Omega $ at small $\Omega $ \cite%
{JC2}.

\subsection{Spontaneous symmetry breaking of the 2D discrete solitons in
linearly-coupled lattices}

A characteristic feature of many nonlinear \textit{dual-core} systems, built
of two identical linearly-coupled waveguides with intrinsic self-attractive
nonlinearity, is a \textit{spontaneous-symmetry-breaking }(SSB)\textit{\ }%
bifurcation, which destabilizes the symmetric ground state, with equal
components of the wave function in the coupled cores, and creates stable
asymmetric states. The SSB bifurcation takes place at a critical value of
the nonlinearity strength, the asymmetric state existing above this value
\cite{SSB}. A system of linearly-coupled DNLS equations is a basic model for
SSB in discrete settings. Its 2D form is \cite{Herring}
\begin{eqnarray}
i\dot{\phi}_{m,n} &=&-(1/4)\left( \phi _{m+1,n}+\phi _{m-1,n}+\phi
_{m,n+1}+\phi _{m,n-1}-4\phi _{m,n}\right)   \notag \\
&&-\left\vert \phi _{m,n}\right\vert ^{2}\phi _{m,n}-K\psi _{m,n}~,  \notag
\\
i\dot{\psi}_{m,n} &=&-(1/4)\left( \psi _{m+1,n}+\psi _{m-1,n}+\psi
_{m,n+1}+\psi _{m,n-1}-4\psi _{m,n}\right)   \notag \\
&&-\left\vert \psi _{m,n}\right\vert ^{2}\psi _{m,n}-K\phi _{m,n}~
\label{coupled}
\end{eqnarray}
where $\phi _{m,n}$ and $\psi _{m,n}$ are wave functions of the discrete
coordinates $m$ and $n$,\ and $K>0$ represents the linear coupling between
the cores. Stationary states with frequency $\omega $ are looked for as $%
\left( \phi _{m,n},\psi _{m,n}\right) =\exp \left( -i\omega t\right) \left(
u_{m,n},v_{m,n}\right) $. Real stationary fields in the two components are
characterized by their norms,%
\begin{equation}
E_{u,v}=\sum_{m,n=-\infty }^{+\infty }\left( u_{m,n}^{2},v_{m,n}^{2}\right) ,
\label{E}
\end{equation}%
and the asymmetry of the symmetry-broken state is determined by parameter%
\begin{equation}
r=\left( E_{u}-E_{v}\right) /\left( E_{u}+E_{v}\right) .  \label{r}
\end{equation}

\begin{figure}
\begin{tabular}{cc}
\includegraphics[width=5.5cm,scale=1]{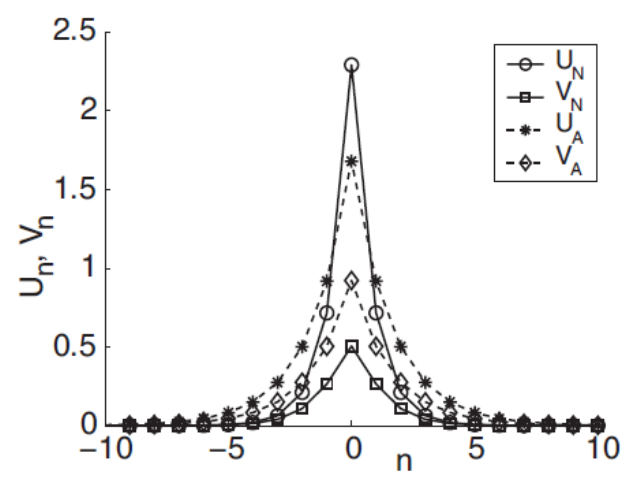} & %
\includegraphics[width=5.5cm,scale=1]{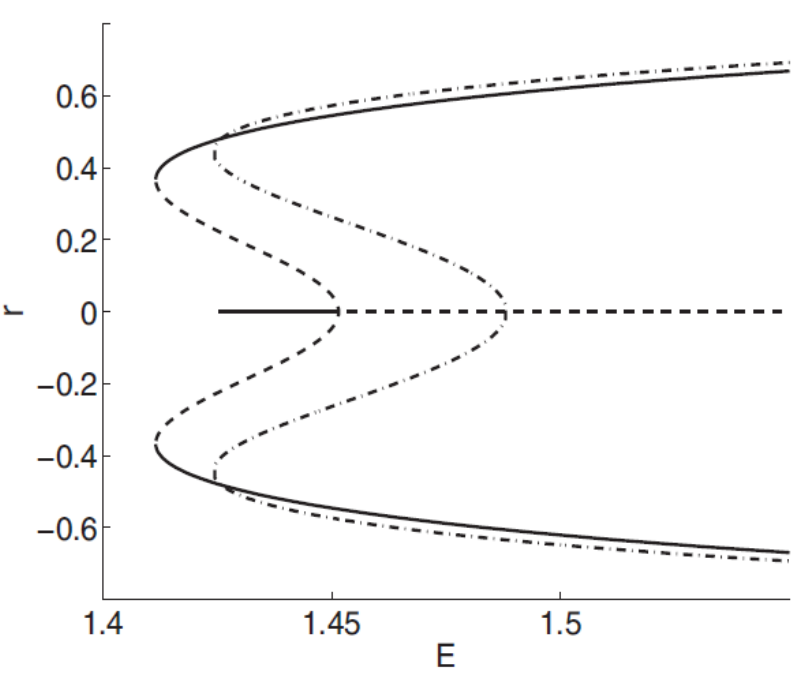}%
\end{tabular}%
\caption{Left: A stable 2D two-component discrete soliton with spontaneously
broken symmetry between the components, generated by system (\protect\ref%
{coupled}). The 2D soliton, with total norm $E\equiv E_{u}+E_{v}=1.435$, is
displayed by means of its cross section. Symbols labelled $\left( U_{\mathrm{%
N}},V_{\mathrm{N}}\right) $ and $\left( U_{\mathrm{A}},V_{\mathrm{A}}\right)
$ stand, respectively, for the components of the numerically constructed
soliton and its analytical counterpart predicted by the VA based on ansatz (%
\protect\ref{SSB-ansatz}). Right: Families of 2D onsite-centered discrete
solitons, generated by system (\protect\ref{coupled}), are shown by means of
curves $r(E)$, where $r$ is the asymmetry parameter (\protect\ref{r}). The
dashed-dotted curve shows the VA prediction, while the solid and dashed ones
represent stable and unstable solitons produced by the numerical solution.
The figure is borrowed from Ref. \protect\cite{Herring}.}
\label{fig4}
\end{figure}

The system under the consideration can be analyzed by means of the VA, based
on the 2D ansatz%
\begin{equation}
\left( u_{m,n},v_{m,n}\right) =\left( A,B\right) \exp \left[ -a\left(
|m|+|n|\right) \right] ,  \label{SSB-ansatz}
\end{equation}%
with inverse width $a$ and amplitudes, $A$ and $B$, of the two components
(cf. the 1D ansatz (\ref{ansatz-onsite})). The SSB is represented by
solutions with $A\neq B$. An example of a stable 2D discrete soliton is
displayed in Fig. \ref{fig4}(a), which corroborates accuracy of the VA. In
Fig. \ref{fig4}(b), the families of symmetric and asymmetric 2D discrete
solitons is characterized by the dependence of asymmetry parameter $r$,
defined as per Eq. (\ref{r}), on the total norm, $E\equiv E_{u}+E_{v}$, see
Eq. (\ref{E}). Figure \ref{fig4}(b) demonstrates the SSB bifurcation of the
\textit{subcritical} type \cite{Iooss}, with the two branches of
broken-symmetry states originally going backward in the $E$ direction, as
unstable ones; they become stable after passing the turning point.
Accordingly,\ Fig. \ref{fig4}(b) demonstrates a bistability area, where
symmetric and asymmetric states coexist as stable ones.

\section{2D discrete solitons in the Salerno model (SM)}

The 2D version of the SM was introduced in Ref. \cite{Zaragoza2D}:
\begin{eqnarray}
i\dot{\psi}_{n,m} &=&-\left[ \left( \psi _{n+1,m}+\psi _{n-1,m}\right)
+\left( \psi _{n,m+1}+\psi _{n,m-1}\right) \right]  \notag \\
&\times &\left( 1+\mu \left\vert \psi _{n,m}\right\vert ^{2}\right)
-2\left\vert \psi _{n,m}\right\vert ^{2}\psi _{n,m}\;,  \label{2dSalerno}
\end{eqnarray}%
Similar to its 1D version (\ref{SAmodel}), Eq. (\ref{2dSalerno}) conserves
the norm and Hamiltonian, cf. Eqs. (\ref{eq:Norm}) and (\ref{HSM}),
\begin{equation}
\left( N_{\mathrm{SM}}\right) _{\mathrm{2D}}=(1/\mu )\sum_{m,n}\ln
\left\vert 1+\mu |\psi _{n,m}|^{2}\right\vert \;,  \label{SalernoNorm}
\end{equation}%
\begin{gather}
\left( H_{\mathrm{SM}}\right) _{\mathrm{2D}}=-\sum_{n,m}\left[ \left( \psi
_{n,m}\psi _{n+1,m}^{\ast }+\psi _{n+1,m}\psi _{n,m}^{\ast }\right) +\left(
\psi _{n,m}\psi _{n,m+1}^{\ast }+\psi _{n,m+1}\psi _{n,m}^{\ast }\right)
\right.  \notag \\
\left. +(2/\mu )|\psi _{n,m}|^{2}\right] +(2/\mu )\left( N_{\mathrm{SM}%
}\right) _{\mathrm{2D}}.\;  \label{eq:SalernoHam}
\end{gather}%
The continuum limit of this model is the 2D equation which is an extension
of its 1D counterpart (\ref{SAcont}):%
\begin{equation}
i\Psi _{t}+\left( 1+\mu \left\vert \Psi \right\vert ^{2}\right) \left( \Psi
_{xx}+\Psi _{yy}\right) +2\left( 2\mu +1\right) |\Psi |^{2}\Psi =0.
\label{cont}
\end{equation}%
Note that the effective nonlinear-diffraction term $\mu \left\vert \Psi
\right\vert ^{2}\left( \Psi _{xx}+\Psi _{yy}\right) $ in Eq. (\ref{cont})
prevents the onset of the collapse because, in the limit of the catastrophic
self-compression, this term becomes dominant, giving a positive contribution
to the energy. Thus, this term makes it possible to construct stable 2D
solitons \cite{Zaragoza2D}.

\begin{figure}
\begin{center}
\includegraphics[width=13cm,scale=1,clip]{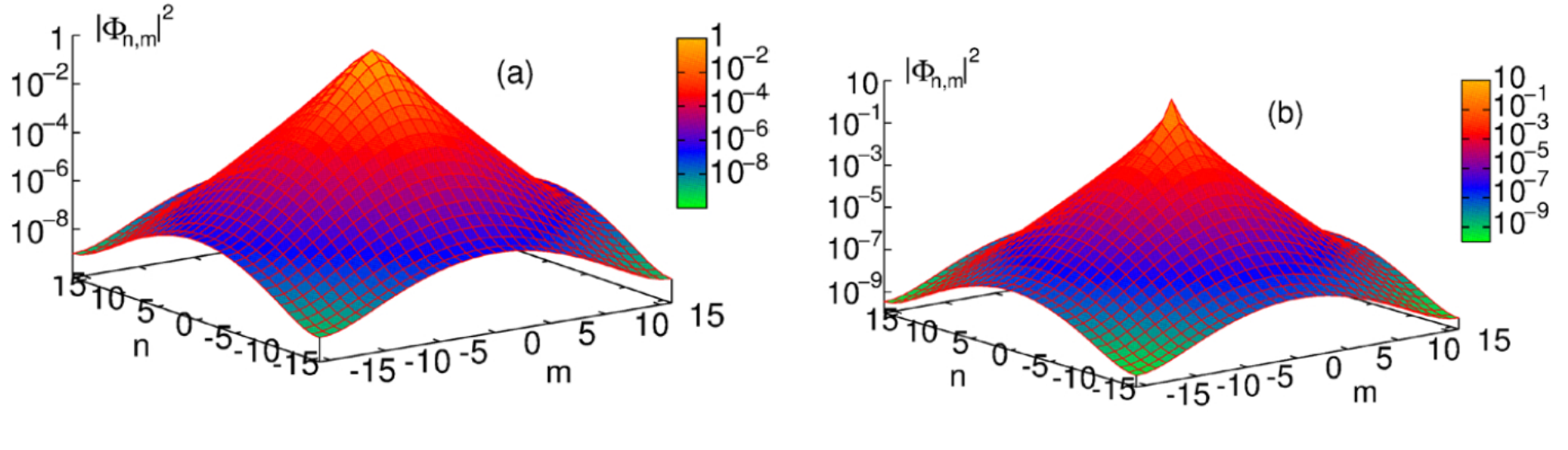}
\end{center}
\caption{Stable discrete solitons in the 2D Salerno model with competing
nonlinearities [$\protect\mu <0$ in Eq. (\protect\ref{2dSalerno})], obtained
for $\protect\omega =-4.22$: (a) a regular soliton at $\protect\mu =-0.2$;
\textbf{(}b\textbf{)} a cuspon at $\protect\mu =-0.88$. The figure is
borrowed from Ref. \protect\cite{Zaragoza2D}.}
\label{fig5}
\end{figure}

2D discrete solitons are looked for as solutions to Eq. (\ref{cont}) in the
usual form, $\psi _{m,n}(t)=e^{-i\omega t}\Phi _{m,n}$. In the most
interesting case of the competing nonlinearities, $\mu <0$, the situation is
similar to that outlined above for the one-dimensional SM: there are
ordinary discrete solitons, which have their stability and instability
regions, and 2D cuspons, which are entirely stable in their existence
region. Typical 2D solitons of both types are displayed in Fig.~\ref{fig5}.
Antisymmetric bound states of ordinary 2D discrete solitons, and symmetric
complexes built of 2D cuspons, are stable, while the bound states of cuspons
with opposite parities are unstable, also like in the 1D model.

\begin{figure}
\begin{center}
\includegraphics[width=9cm,scale=1,clip]{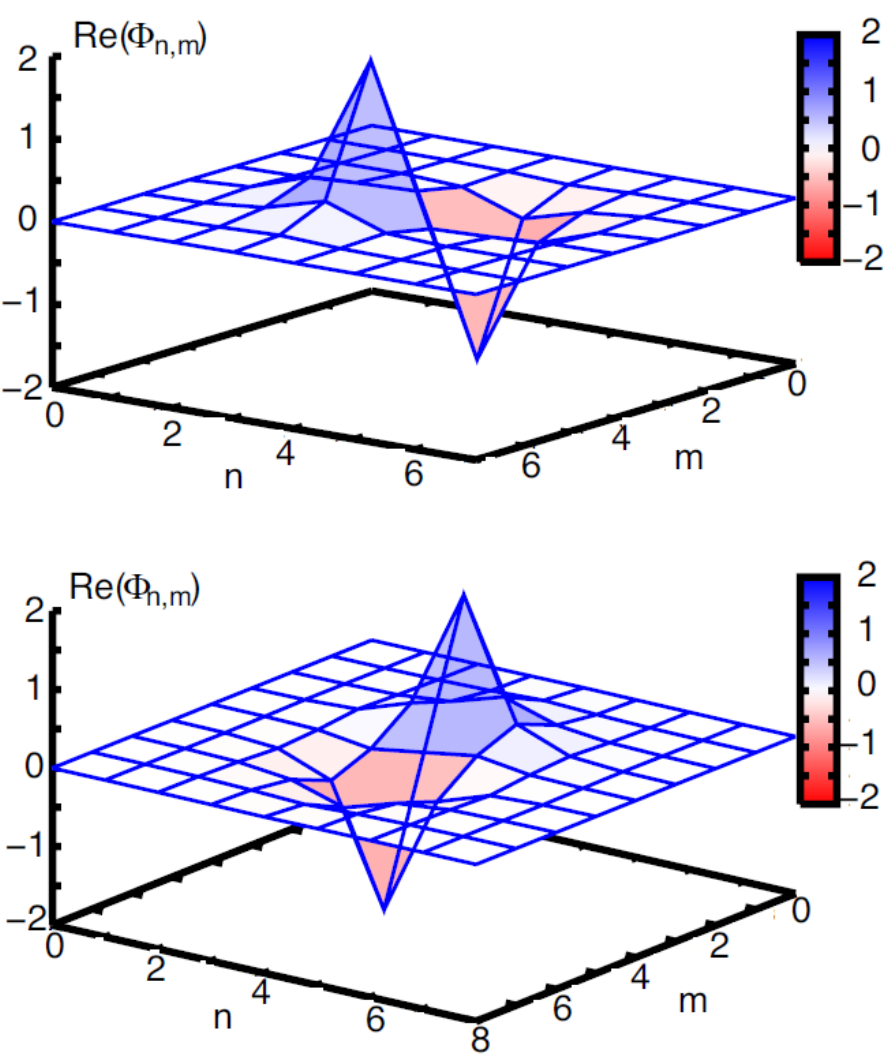}
\end{center}
\caption{Examples of discrete vortex solitons with topological charge $S=1$,
produced by the 2D SM, based on Eq. (\protect\ref{2dSalerno}). Profiles of
the real part of the stationary wave function $\Phi _{m,n}$ for the vortices
of the onsite-centered (stable \textit{vortex cross}) and intersite-centered
(unstable \textit{vortex square}) types are displayed in the top and bottom
panels, respectively. Both solutions are obtained for $\protect\mu =-0.4$
and $\protect\omega =7.0$. The figure is borrowed from Ref. \protect\cite%
{Zaragoza2D}.}
\label{fig6}
\end{figure}

Along with the fundamental solitons, the 2D SM with the competing
nonlinearities gives rise to vortex-soliton modes which may be stable in
narrow parameter regions \cite{Zaragoza2D}. Examples of onsite- and
intersite-centered vortex solitons (alias \textit{vortex cross} and \textit{%
vortex square}, respectively) are presented in Fig. \ref{fig6}. In the 2D SM
with non-competing nonlinearities ($\mu >0$ in Eq. (\ref{cont})), vortex
solitons are unstable, spontaneously transforming into fundamental ones and
losing their vorticity. This transition is possible because the angular
momentum is not conserved in the lattice system. The situation is different
in the 2D SM with competing nonlinearities ($\mu <0$), where unstable vortex
modes transform into \textit{vortical breathers}, i.e., persistently
oscillating localized modes that keep the original vorticity.

\section{Solitons of the semi-vortex (SV) and mixed-mode (MM) types in the
discrete 2D spin-orbit-coupling (SOC) system}

Recently, much interest has been drawn to emulation of the
spin-orbit-coupling (SOC) effect, which is well known in physics of
semiconductors, in binary BEC \cite{Spielman,Galitski,Goldman,Zhai}. While
SOC is a linear effect, its interplay with the intrinsic mean-field
nonlinearity of atomic BEC gives rise to predictions of new species of 1D,
2D, and 3D solitons \cite{EPL}. In particular, the effectively 2D binary BEC
with SOC of the \textit{Rashba type} is modeled by the following system of
coupled GP equations for two components $\phi ^{(\pm )}$ of the
pseudo-spinor wave function \cite{Ben Li},
\begin{eqnarray}
i\frac{\partial \phi ^{(+)}}{\partial t} &=&-\frac{1}{2}\nabla ^{2}\phi
^{(+)}-(|\phi ^{(+)}|^{2}+\gamma |\phi ^{(-)}|^{2})\phi ^{(+)}+2\lambda
\left( \frac{\partial \phi ^{(-)}}{\partial x}-i\frac{\partial \phi ^{(-)}}{%
\partial y}\right) ,  \notag \\
i\frac{\partial \phi ^{(-)}}{\partial t} &=&-\frac{1}{2}\nabla ^{2}\phi
^{(-)}-(|\phi ^{(-)}|^{2}+\gamma |\phi ^{(+)}|^{2})\phi ^{(-)}+2\lambda
\left( -\frac{\partial \phi ^{(+)}}{\partial x}-i\frac{\partial \phi ^{(+)}}{%
\partial y}\right) .  \label{Rashba}
\end{eqnarray}%
In this system, SOC of the \textit{Rashba type} is represented in by terms
with coefficient $\lambda $, which couple the two equations through the
first-order spatial derivatives. The system also includes the self- and
cross-attractive nonlinearities, with scaled coefficients $1$ and $\gamma $,
respectively.

The system of coupled GP equations (\ref{Rashba}) maintains 2D solitons of
two different types, namely, semi-vortices (SVs) and mixed modes (MMs) \cite%
{Ben Li}, The SV solitons, written in polar coordinates $\left( r,\theta
\right) $, have zero vorticity in one component, and vorticity $+1$ or $-1$
in the other:%
\begin{eqnarray}
\phi _{1}^{(+)} &=&e^{-i\omega t}f_{1}(r^{2}),~\phi _{1}^{(-)}=e^{-i\omega
t+i\theta }rf_{2}(r^{2}),  \label{01} \\
\phi _{2}^{(+)} &=&e^{-i\omega t-i\theta }rf_{2}(r^{2}),~\phi
_{2}^{(-)}=e^{-i\omega t}f_{1}(r^{2}),  \label{10}
\end{eqnarray}%
where $\omega $ is the chemical potential, and $f_{1,2}\left( r^{2}\right) $
are real functions which take finite values at $r=0$ and exponentially decay
$\sim \left( \sin \left( 2\lambda r\right) ,\cos \left( 2\lambda r\right)
\right) \exp \left( -\sqrt{2\left( -\omega -2\lambda ^{2}\right) }r\right) $
at $r\rightarrow \infty $. The two SV solutions (\ref{01}) and (\ref{10}),
which are mirror images of each other, exist in the semi-infinite bandgap, $%
\omega <-2\lambda ^{2}$.

The combination of zero and nonzero vorticities in the SV solutions (\ref{01}%
) and (\ref{10}) is exactly compatible with the structure of the coupled GP
equations (\ref{Rashba}). On the contrary to this, MM solitons cannot be
represented by an exact ansatz similar to Eqs. (\ref{01}) and (\ref{10}),
but they may be approximated by a linear combination of both types of the
SVs, $\left\{ \phi _{1}^{(+)}+\phi _{2}^{(+)},\phi _{1}^{(-)(1)}+\phi
_{2}^{(+)}\right\} $. An essential result is that the SVs and MMs are stable
and represent the system's ground state in the cases of $\gamma <1$ and $%
\gamma >1$, respectively, i.e., when the self-attraction is stronger or
weaker than the cross-attraction in Eqs. (\ref{Rashba}) \cite{Ben Li}. On
the other hand, the SVs and MMs are unstable, as excited states, in the
opposite cases, i.e., $\gamma >1$ and $\gamma <1$, respectively.

The discretized version of the SOC GP system (\ref{Rashba}), which
corresponds to the spin-orbit-coupled binary BEC trapped in a deep OL
potential, with discrete coordinates $(m,n)$, was introduced in Ref. \cite%
{HS}:
\begin{gather}
i\frac{\partial \phi _{m,n}^{(+)}}{\partial t}=-\frac{1}{2}\left( \phi
_{m+1,n}^{(+)}+\phi _{m-1,n}^{(+)}+\phi _{m,n+1}^{(+)}+\phi
_{m,n-1}^{(+)}-4\phi _{m,n}^{(+)}\right)  \notag \\
-(|\phi _{m,n}^{(+)}|^{2}+\gamma |\phi _{-}|^{2})\phi _{m,n}^{(+)}+\lambda
\left[ \phi _{m+1,n}^{(-)}-\phi _{m-1,n}^{(-)}-i\left( \phi
_{m,n+1}^{(-)}-\phi _{m,n-1}^{(-)}\right) \right] ,  \notag \\
i\frac{\partial \phi _{m,n}^{(-)}}{\partial t}=-\frac{1}{2}\left( \phi
_{m+1,n}^{(-)}+\phi _{m-1,n}^{(-)}+\phi _{m,n+1}^{(-)}+\phi
_{m,n-1}^{(-)}-4\phi _{m,n}^{(-)}\right)  \notag \\
-(|\phi _{m,n}^{(-)}|^{2}+\gamma |\phi _{m,n}^{(+)}|^{2})\phi _{-}+\lambda
\left[ -\left( \phi _{m+1,n}^{(+)}-\phi _{m-1,n}^{(+)}\right) -i\left( \phi
_{m,n+1}^{(+)}-\phi _{m,n-1}^{(+)}\right) \right] .  \label{2D-discr}
\end{gather}%
The linearized version of this system gives rise to the following dispersion
relation for the discrete plane waves, $\phi _{m,n}^{(\pm )}\sim \exp \left( -i\omega
t+ipx+iqy\right) $, with wavenumbers \ taking values in the first Brillouin
zone, $0<p,q<2\pi $:%
\begin{equation}
\omega =2\left( \sin ^{2}\frac{p}{2}+\sin ^{2}\frac{q}{2}\right) \pm
2\lambda \sqrt{\sin ^{2}p+\sin ^{2}q}.  \label{spectrum}
\end{equation}

\begin{figure}
\begin{tabular}{cc}
\includegraphics[width=5.5cm,scale=1]{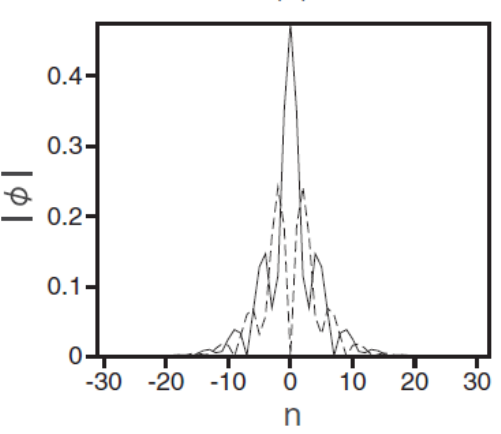} & %
\includegraphics[width=5.5cm,scale=1]{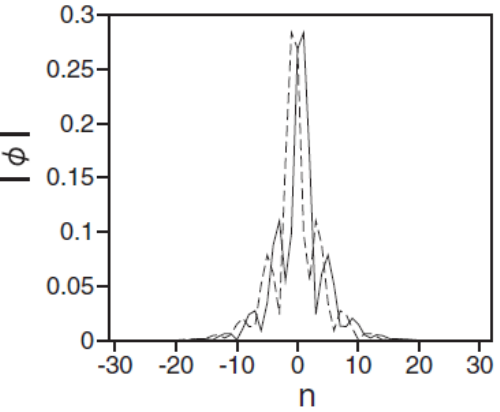}%
\end{tabular}%
\caption{Left: Juxtaposed profiles of $\left\vert \protect\phi %
_{m,n}^{(+)}\right\vert $ and $\left\vert \protect\phi _{m,n}^{(-)}\right%
\vert $ (solid and dashed lines, respectively) of a stable 2D discrete
soliton of the semi-vortex type, in the central cross section, produced by a
numerical solution of Eq. (\protect\ref{2D-discr}) with $\protect\lambda %
=0.53$ and $\protect\gamma =0$. The soliton's norm (see Eq. (\protect\ref{N}%
)) is $N=3.5$. Right: The same for a stable discrete soliton of the
mixed-mode type, with $\protect\lambda =0.58$, $\protect\gamma =2$, and $N=2$%
. Values of the discrete fileds at lattice sites are connected by lines, for
better visualization. The figure is borrowed from Ref. \protect\cite{HS}.}
\label{fig10}
\end{figure}

The numerical solution of Eq. (\ref{2D-discr}) has produced 2D modes which
are discrete counterparts of the SV and MM solitons of the continuum system (%
\ref{Rashba}), see examples in Fig. \ref{fig10}. As concerns the stability,
the discreteness extends the stability of the SV and MM\ solitons towards $%
\gamma >1$ and $\gamma <1$, respectively.

\begin{figure}
\begin{tabular}{cc}
\includegraphics[width=5.5cm,scale=1]{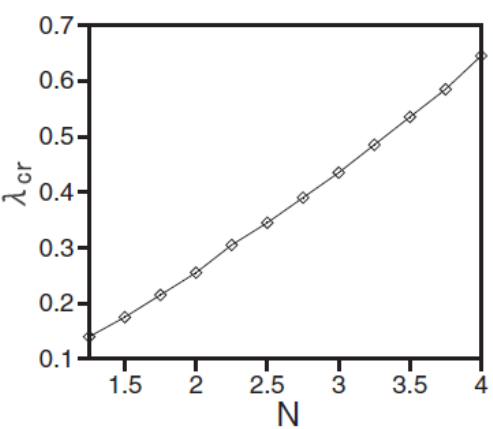} & %
\includegraphics[width=5.5cm,scale=1]{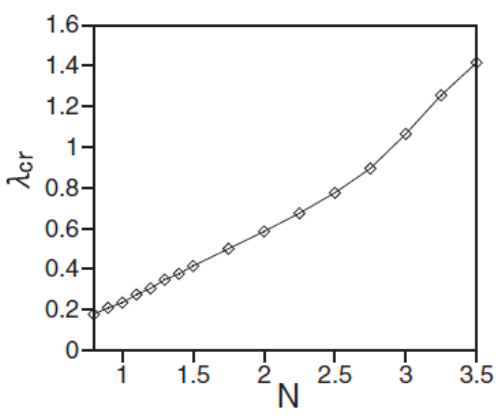}%
\end{tabular}%
\caption{Left and right: The dependence of the critical value of the SOC\
strength, $\protect\lambda _{\mathrm{cr}}$, above which the 2D discrete
solitons of the SV and MM types, produced by the numerical solution of Eq. (%
\protect\ref{2D-discr}) with $\protect\gamma =0$ and $2$, respectively,
suffer the delocalization, on the total soliton's norm (\protect\ref{N}).
The figure is borrowed from Ref. \protect\cite{HS}.}
\label{fig10additional}
\end{figure}

A drastic difference of the discrete solitons of both the SV and MM types
from their counterparts in the continuum is that they suddenly suffer
delocalization (decay) when the SOC strength $\lambda $ in Eq. (\ref%
{2D-discr}) exceeds a certain critical value, $\lambda _{\mathrm{cr}}$. The
dependence of $\lambda _{\mathrm{cr}}$ on the soliton's norm,
\begin{equation}
N=\sum_{m,n}\left( \left\vert \phi _{m,n}^{(+)}\right\vert ^{2}+\left\vert
\phi _{m,n}^{(-)}\right\vert ^{2}\right) ,  \label{N}
\end{equation}%
for the SV and MM solitons is displayed in Fig. \ref{fig10additional}. The
onset of the delocalization may be explained as a transition of the solution
from the spectral bandgap to the band populated by the small-amplitude
plane-wave states in the system's linear spectrum, which is produced by Eq. (%
\ref{spectrum}).

\section{Stable soliton species in the 3D DNLS equation}

\subsection{The 3D setting}

A natural development of the analysis of the solitons and solitary vortices,
and their bound states, produced by the 2D discrete DNLS equation and its
extensions, which is outlined above in Sections 2 -- 4, is to construct
self-trapped states (solitons) in the framework of the 3D equation:%
\begin{gather}
i\dot{\phi}_{l,m,n}+C\left( \phi _{l+1,m,n}+\phi _{l,m+1,n}+\phi
_{l,m,n+1}+\phi _{l-1,m,n}+\phi _{l,m-1,n-1}+\phi _{l,m,n-1}-6\phi
_{l,m,n}\right)  \notag \\
+\left\vert \phi _{l,m,n}\right\vert ^{2}\phi _{l,m,n}=0,  \label{3D}
\end{gather}%
where, as above, the overdot stands for the time derivative, $\left(
l,m,n\right) $ is the set of the 3D discrete coordinates, and $C>0$ is the
coefficient of the intersite coupling. The 3D DNLS equation cannot be
realized in optics, but it admits natural implementation for BEC loaded in a
deep 3D OL potential \cite{Brazhnyi,Morsch}. In that case, $\phi _{l,m,n}(t)$
is the the respective effectively discretized BEC wave function.

As above, stationary soliton solutions to Eq. (\ref{3D}) with chemical
potential $\omega $ are looked for as
\begin{equation}
\phi _{l,m,n}=\exp (-i\omega t)u_{l,m,n},  \label{omega}
\end{equation}%
where the stationary discrete wave function $u_{l,m,n}$ obeys the
corresponding equation,
\begin{gather}
\omega u_{l,m,n}+C\left(
u_{l+1,m,n}+u_{l,m+1,n}+u_{l,m,n+1}+u_{l-1,m,n}+u_{l,m-1,n-1}+u_{l,m,n-1}-6u_{l,m,n}\right)
\notag \\
+\left\vert u_{l,m,n}\right\vert ^{2}u_{l,m,n}=0.  \label{standing}
\end{gather}%
In particular, numerical solutions of Eq. (\ref{standing}) for 3D discrete
solitons with embedded vorticity $S=0,1,2,...$ ($S=0$ corresponds to the
fundamental solitons, for which the wave function $u_{l,m,n}$ is real) can
be obtained, starting from the natural input
\begin{equation}
u_{l,m,n}^{(\mathrm{init})}=A\left( l+im\right) ^{S}\mathrm{sech}\left( \eta
\sqrt{l^{2}+m^{2}}\right) \exp \left( -|n|\right) ,  \label{3Dansatz}
\end{equation}%
where $\eta $ is a real scale parameter, and it is implied that the
vorticity axis is directed along coordinate $n$ \cite{3Dsol1}.

It is also relevant to consider a two-component system of
nonlinearly-coupled 3D DNLS equations, for wave functions $\phi _{l,m,n}(t)$
and $\psi _{l,m,n}(t)$ of two interacting BEC\ species (most typically,
these are different hyperfine states of the same atom) \cite{3Dsol1}:

\begin{gather}
i\dot{\phi}_{l,m,n}+C\left( \phi _{l+1,m,n}+\phi _{l,m+1,n}+\phi
_{l,m,n+1}+\phi _{l-1,m,n}+\phi _{l,m-1,n-1}+\phi _{l,m,n-1}-6\phi
_{l,m,n}\right)  \notag \\
+\left( \left\vert \phi _{l,m,n}\right\vert ^{2}+\beta |\psi
_{l,m,n}|^{2}\right) \phi _{l,m,n}=0,  \notag \\
i\dot{\psi}_{l,m,n}+C\left( \psi _{l+1,m,n}+\psi _{l,m+1,n}+\psi
_{l,m,n+1}+\psi _{l-1,m,n}+\psi _{l,m-1,n-1}+\psi _{l,m,n-1}-6\psi
_{l,m,n}\right)  \notag \\
+\left( \left\vert \psi _{l,m,n}\right\vert ^{2}+\beta |\phi
_{l,m,n}|^{2}\right) \psi _{l,m,n}=0,  \label{3D system}
\end{gather}%
Here $\beta >0$ is the relative strength of the inter-component attractive
interaction with respect to the intra-component self-attraction.

\subsection{Results}

\subsubsection{Single-component 3D solitons}

The numerical analysis, starting from input (\ref{3Dansatz}), has provided
families of fundamental and vortex solitons. Here, following Ref. \cite%
{3Dsol1}, the results are displayed for a fixed value of the chemical
potential, $\omega =-2$ in Eq. (\ref{omega}), while varying coupling
constant $C$ in Eqs. (\ref{3D}) and (\ref{standing}). In particular, the
discrete fundamental solitons with $S=0$ are stable at $C<$ $C_{\mathrm{cr}%
}^{(0)}\approx 2$, and the vortex modes with $S=1$ are stable at $C<$ $C_{%
\mathrm{cr}}^{(1)}\approx 0.65$. Note that the limit of $C\rightarrow \infty
$ corresponds to the 3D NLS equation in the continuum, in which all solitons
are definitely unstable; therefore, all discrete solitons become unstable at
sufficiently large values of $C$. At $C>C_{\mathrm{cr}}^{(1)}$ the
simulations demonstrate that the development of the instability destroys the
vortical structure and, eventually, transforms the soliton into a
fundamental one, with $S=0$ (not shown here in detail). The change of the
topological charge is possible, as the angular momentum is not a dynamical
invariant of the lattice dynamics.

\begin{figure}
\begin{center}
\includegraphics[width=11cm,scale=1,clip]{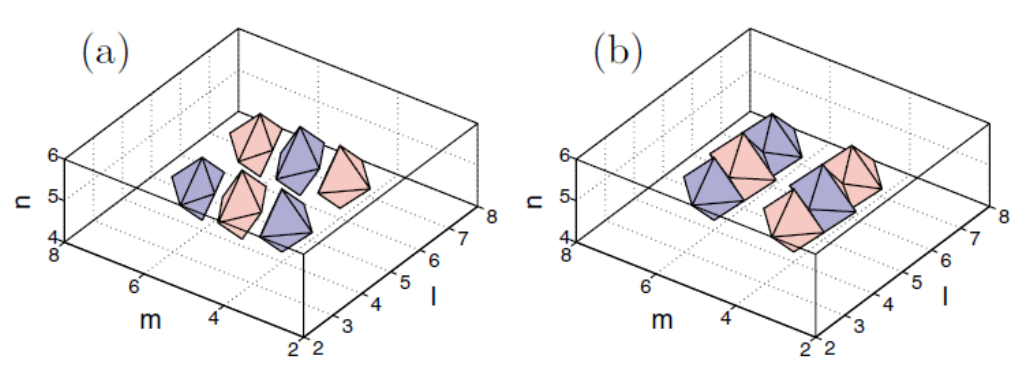}
\end{center}
\caption{A stable discrete vortex soliton with topological charge $S=3,$
produced by the numerical solution of Eq. (\protect\ref{standing}) with $%
C=0.01$ and $\protect\omega =-2$. Panels (a) and (d) display level contours
corresponding to Re$\left( u_{l,m,n}\right) =\pm 0.25$ and Im$\left(
u_{l,m,n}\right) =\pm 0.25$, respectively.The figure is borrowed from Ref.
\protect\cite{3Dsol1}.}
\label{fig14}
\end{figure}

The vortex solitons with $S=2$ are completely unstable, but an unusual
feature of these states is that, at sufficiently small values of $C$ (in
particular, at $C=0.01$), the instability spontaneously rebuilds them into
stable discrete solitons with a \emph{larger vorticity}, $S=3$ \cite{3Dsol1}%
. An example of a stable soliton with $S=3$ is displayed in Fig. \ref{fig14}.

\begin{figure}
\begin{center}
\includegraphics[width=11cm,scale=1,clip]{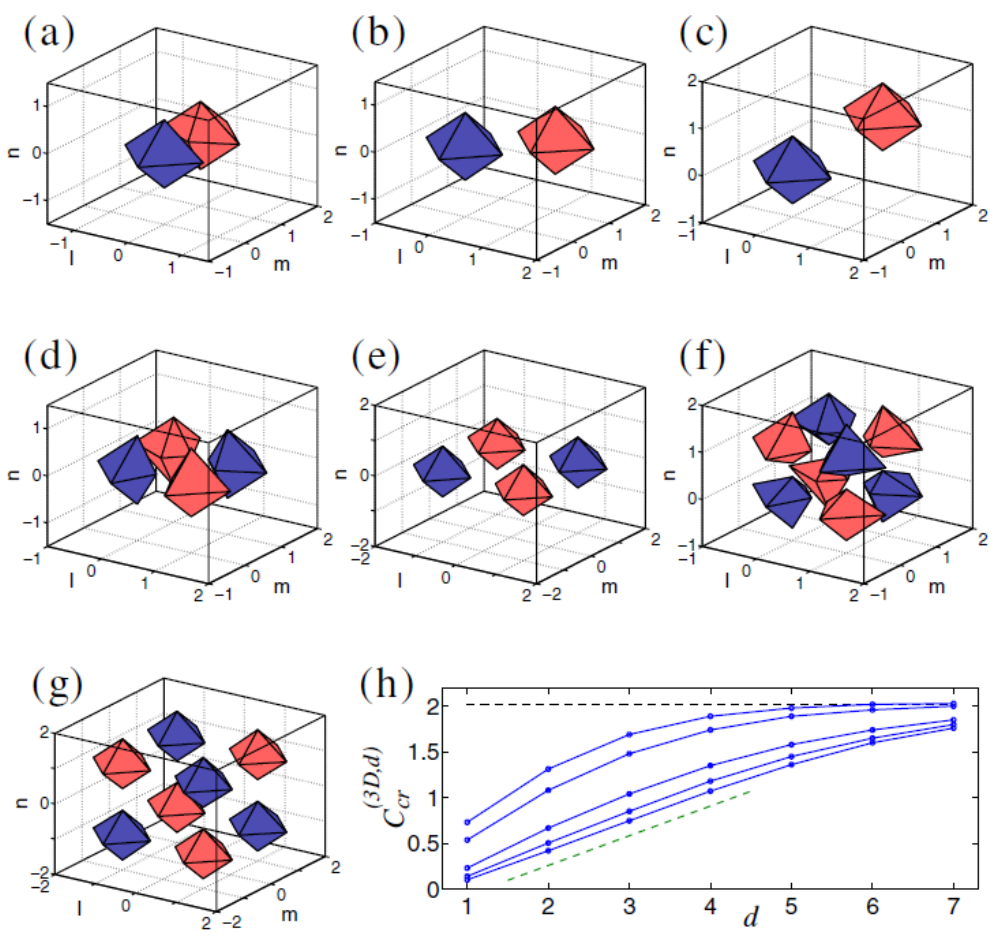}
\end{center}
\caption{Stable 3D multipole solutions of Eq. (\protect\ref{standing}) with $%
C=0.1$ and $\protect\omega =-2$. The top row depicts stable tightly-bound
dipoles (with intersite separation $d=1$): (a) straight, (b) 2D-diagonal,
and (c) 3D-diagonal ones. (d) and (e): Quadrupoles set in the $n=0$ plane,
with intrinsic separation $d=1$ and $d=2$, respectively. (f) and (g):
Octupoles with $d=1$ and $d=2$. Panel (h) displays the stability boundary $%
C_{cr}^{\mathrm{(}3D\mathrm{,}d\mathrm{)}}$ as a function of the intrinsic
separation $d$ for diagonal, oblique, and straight dipoles, octupoles, and
quadrupoles, from top to bottom. The horizontal dashed line designates the
stability threshold for the fundamental discrete soliton. Note that, for the
quadrupoles (the bottom boundary), $C_{\mathrm{quad}}^{\mathrm{(}3D\mathrm{,}%
d\mathrm{)}}$ is a linear function of $d$ at $d\leq 4$ (see the dashed
straight line with slope $0.325$, included for the guidance). In panels
(a)-(g), level contour corresponding to $\mathrm{Re}(u_{l,m,n})=\pm 0.5$ are
shown by blue and red (colors, respectively. The figure is borrowed from
Ref. \protect\cite{3Dsol2}.}
\label{fig15}
\end{figure}

In addition to the fundamental and vortex solitons, Eqs. (\ref{3D}) and (\ref%
{standing}) produce diverse multimode species of stable discrete 3D solitons
in the form of dipoles, quadrupoles and octupoles \cite{3Dsol2}. Examples of
such states are presented in Fig. \ref{fig15} for $C=0.1$. This figure
displays tightly-bound dipoles with different orientations with respect to
the lattice, $viz$., straight, 2D-diagonal, and 3D-diagonal ones (panels
(a,b,c)), quadrupole (panel (d)), and octupole (panel (f)), in which the
field fills adjacent sites of the lattice (with the lattice distance between
them $d=1$). Also displayed are loosely-bound quadrupole and octupole (in
panels (e) and (g), respectively), with distance $d=2$ between the filled
sites. Similar multimode states with still larger separations $d$ between
the filled sites were found too. The results are summarized in Fig. \ref%
{fig15}(h), which shows stability boundaries $C_{cr}^{(3D,d)}$ for dipoles,
quadrupoles, and octupoles vs. $d$. Naturally, the stability region, $%
C<C_{cr}^{(3D,d)}$, increases with the increase of $d$, as the interaction
between the filled sites, which leads to the possible dynamical instability
of the multipole states, is weaker for larger $d$.

\begin{figure}
\begin{center}
\includegraphics[width=11cm,scale=1,clip]{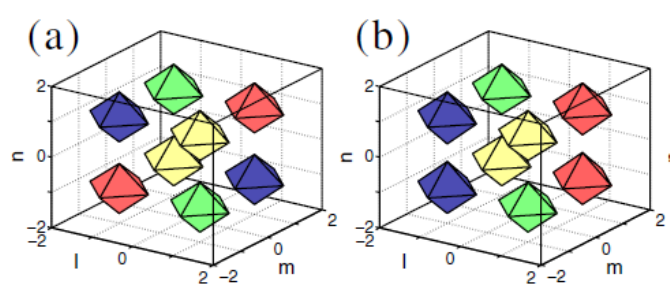}
\end{center}
\caption{Vortex cubes produced by the numerical solution of Eq. (\protect\ref%
{standing}) with $\Lambda =2$ and $C=0.1$. Panel (a) shows a stable
composite mode, built of two parallel identical quasi-planar vortices with
topological numbers $S_{1}=S_{2}=1$ and a phase shift of $\protect\pi $.
Panel (b) shows an unstable vortex-antivortex cube, formed by vortices with
opposite topological charges, $S_{1}=-S_{2}$. Level contours corresponding
to $\mathrm{Re}(u_{l,m,n})=\pm 0.5$ are shown by blue and red colors, and
the contours corresponding to $\mathrm{Im}(u_{l,m,n})=\pm 0.5$ are shown by
green and yellow colors, respectively. The figure is borrowed from Ref.
\protect\cite{3Dsol2}.}
\label{fig16}
\end{figure}

In addition to the above-mentioned states, Eq. (\ref{standing}) admits more
sophisticated stable composite states, such as \textquotedblleft vortex
cubes", built as a pair of identical parallel quasi-planar vortices with
topological numbers $S_{1}=S_{2}=1$, with opposite signs (phase shift $\pi $%
). set in parallel planes, as shown in Fig. \ref{fig16}(a). Stationary
solutions representing vortex-antivortex cubes, in the form of parallel
quasi-planar vortices with opposite topological charges, $S_{1}=-S_{2}=1$,
can be found too, as shown in Fig. \ref{fig16}(b), but they are completely
unstable.

The same Eq. (\ref{3D}) gives rise to other stable self-trapped modes, such
as vortex solitons with the axis directed along the 2D diagonal, cf. Fig. %
\ref{fig15}(b). Vortex modes with the axis parallel to the 3D diagonal exist
too, but they are unstable, see further details in Ref. \cite{3Dsol2}.

\paragraph{Two-component 3D solitons (including skyrmions)}

\begin{figure}
\begin{center}
\includegraphics[width=11cm,scale=1,clip]{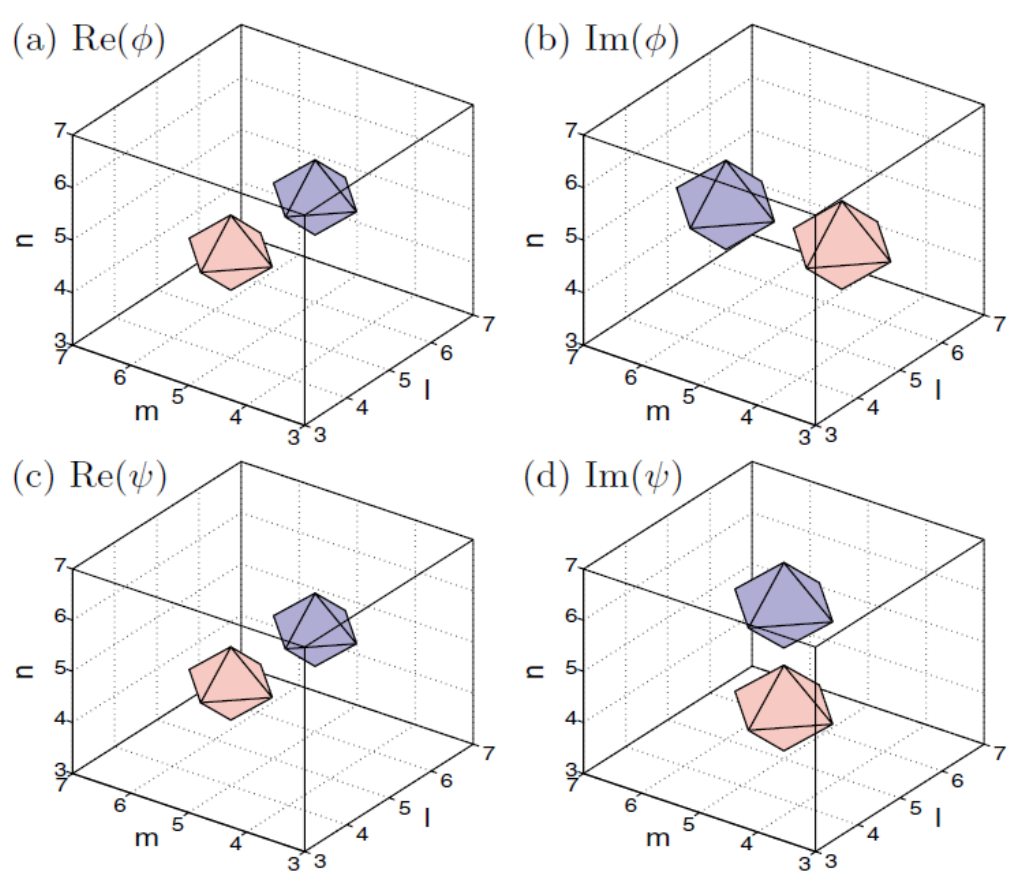}
\end{center}
\caption{A stable bound state of two mutually orthogonal vortex solitons
with topological charges $S=-1$. The bound state is obtained as a numerical
solution of Eqs. (\protect\ref{3D system}) with $\protect\beta =0.5$, $%
C=0.01 $, and chemical potentials of both components $\protect\omega =-2$.
Top and bottom panels represent, respectively, to the first and second
components of the system. The panels show contours of $\mathrm{Re}%
(u_{l,m,n})=\pm 0.5$ and $\mathrm{Im}(u_{l,m,n})=\pm 0.5$. The bluish and
reddish colors designate the positive and negative values, respectively. The
figure is borrowed from Ref. \protect\cite{3Dsol1}.}
\label{fig17}
\end{figure}

The system of coupled 3D DNLS equations (\ref{3D system}) produces stable
soliton states which are specific to the two-component nonlinear lattice
medium. A noteworthy example is a composite mode built as a bound state of
vortex solitons in the two components with \textit{mutually orthogonal}
orientations, see an example in Fig. \ref{fig17}. These bound states are
stable for sufficiently small values of the coupling constant, such as $%
C=0.01$ in Fig. \ref{fig17}, and for $\beta <1$ in Eq. (\ref{3D system}),
i.e., under the condition that the self-attraction in each component is
stronger than the inter-component attraction.

\begin{figure}
\begin{center}
\includegraphics[width=11cm,scale=1,clip]{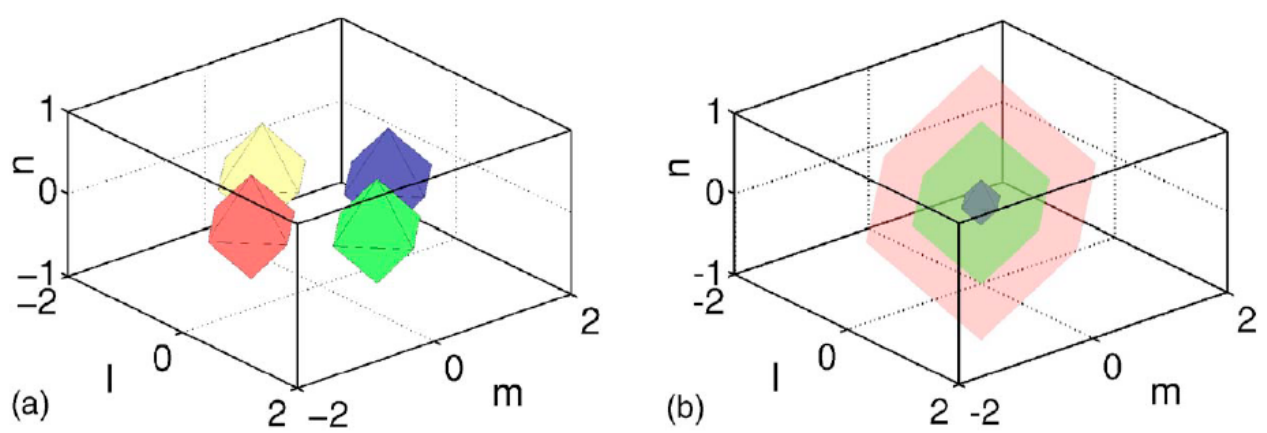}
\end{center}
\caption{An example of a 3D discrete skyrmion, produced by the numerical
solution of Eqs. (\protect\ref{uv}) with $C=0.05$, $\protect\beta =0.25$,
and $\protect\omega =2$. The left and right panels show, respectively,
contours of the complex field, corresponding to $\mathrm{Re}(u)=+1/-1$
(blue/red colors) and $\mathrm{Im}(u)=+1/-1$ (green/yellow colors), and of
the real field, corresponding to $v=(1,0,-1)$ (blue, green and red colors,
designating the contours from the inside out). The figure is borrowed from
Ref. \protect\cite{skyrmion}.}
\label{fig18}
\end{figure}

The system of coupled GP equations with the repulsive sign of the
nonlinearity may be used as the simplest model producing skyrmions in the
binary BEC \cite{skyrm1,skyrm2,skyrm3}. The discretization of the GP system
leads to Eqs. (\ref{3D system}) with the opposite sign in front of the
nonlinear terms \cite{skyrmion}. Then, these equations are reduced to
stationary ones by the usual substitution with chemical potential $\omega $,
$\left\{ \phi ,\psi \right\} =\exp \left( -i\omega t\right) \left\{
u_{l,m,n},v_{l,m,n}\right\} $:%
\begin{gather}
C\left(
u_{l+1,m,n}+u_{l,m+1,n}+u_{l,m,n+1}+u_{l-1,m,n}+u_{l,m-1,n-1}+u_{l,m,n-1}-6u_{l,m,n}\right)
\notag \\
-\left( \left\vert u_{l,m,n}\right\vert ^{2}+\beta |v_{l,m,n}|^{2}\right)
u_{l,m,n}=-\omega u_{l,m,n}~,  \notag \\
C\left(
v_{l+1,m,n}+v_{l,m+1,n}+v_{l,m,n+1}+v_{l-1,m,n}+v_{l,m-1,n-1}+v_{l,m,n-1}-6v_{l,m,n}\right)
\notag \\
-\left( \left\vert v_{l,m,n}\right\vert ^{2}+\beta |u_{l,m,n}|^{2}\right)
v_{l,m,n}=-\omega v_{l,m,n}~,  \label{uv}
\end{gather}%
where the relative strength $\beta $ of the inter-component repulsion with
respect to the self-repulsion remain a positive coefficient. For $\omega >0$%
, skyrmions can be constructed by choosing field $u_{l,m,n}$ as a complex
one, representing a quasi-flat vortex soliton with topological charge $S=1$,
and \textit{real field} $v_{l,m,n}$ as a \textit{bubble} into which the
vortex soliton is embedded, with a nonzero background value at $\left(
|l|,|m|,|n|\right) \rightarrow \infty $, \textit{viz}., $v_{\mathrm{%
background}}^{2}=\omega $ \cite{skyrmion}. An example of a numerically found
skyrmion solution of this type is displayed in Fig. \ref{fig18}.

The same work \cite{skyrmion} reported solutions for 2D discrete
\textquotedblleft baby skyrmions", which are lattice counterparts of the
modes produced by the 2D reduction of the Skyrme model \cite{baby1,baby2}.
The have a simple structure similar to its 3D counterpart displayed in Fig. %
\ref{fig18}, i.e., a complex 2D vortex soliton in one component, embedded
into a bubble of the delocalized field in the other real component.

\section{2D solitons and solitary vortices in semi-discrete systems}

\subsection{Spatiotemporal optical solitons in arrayed waveguides}

The consideration of 2D and 3D settings suggests a natural option to
introduce 2D semi-discrete systems, with a continuous coordinate in one
direction and a discrete coordinate in another, as well as 3D systems, where
one or two coordinates are continuous, while the remaining two or one
coordinates are discrete. In optics, a well-known 2D setting belonging to
this class represents spatiotemporal propagation of light in an array of
optical fibers \cite{Rubenchik}. It is modeled by the system of
linearly-coupled NLS equations for complex amplitudes $u_{n}\left( z,\tau
\right) $ of optical fields in individual fibers:
\begin{equation}
i\partial _{z}u_{n}+(1/2)D\partial _{\tau }^{2}u_{n}+(\kappa /2)\left(
u_{n+1}+u_{n-1}-2u_{n}\right) +\left\vert u_{n}\right\vert ^{2}u_{n}=0,
\label{kappa}
\end{equation}%
where $z$ is the propagation distance, $\tau \equiv t-x/V_{\mathrm{gr}}$
(with time $t$ and carrier group velocity $V_{\mathrm{gr}}$) is the usual
temporal variable, real $D$ is the group-velocity-dispersion coefficient in
each fiber, $\kappa >0$ is the coefficient of coupling between adjacent
fibers in the array, and the nonlinearity coefficient is normalized to be $1$%
. It is commonly known that optical solitons (semi-discrete ones, in the
present case) can be supported in the case of anomalous dispersion, i.e., $%
D>0$.

\begin{figure}
\includegraphics[width=6cm,scale=1]{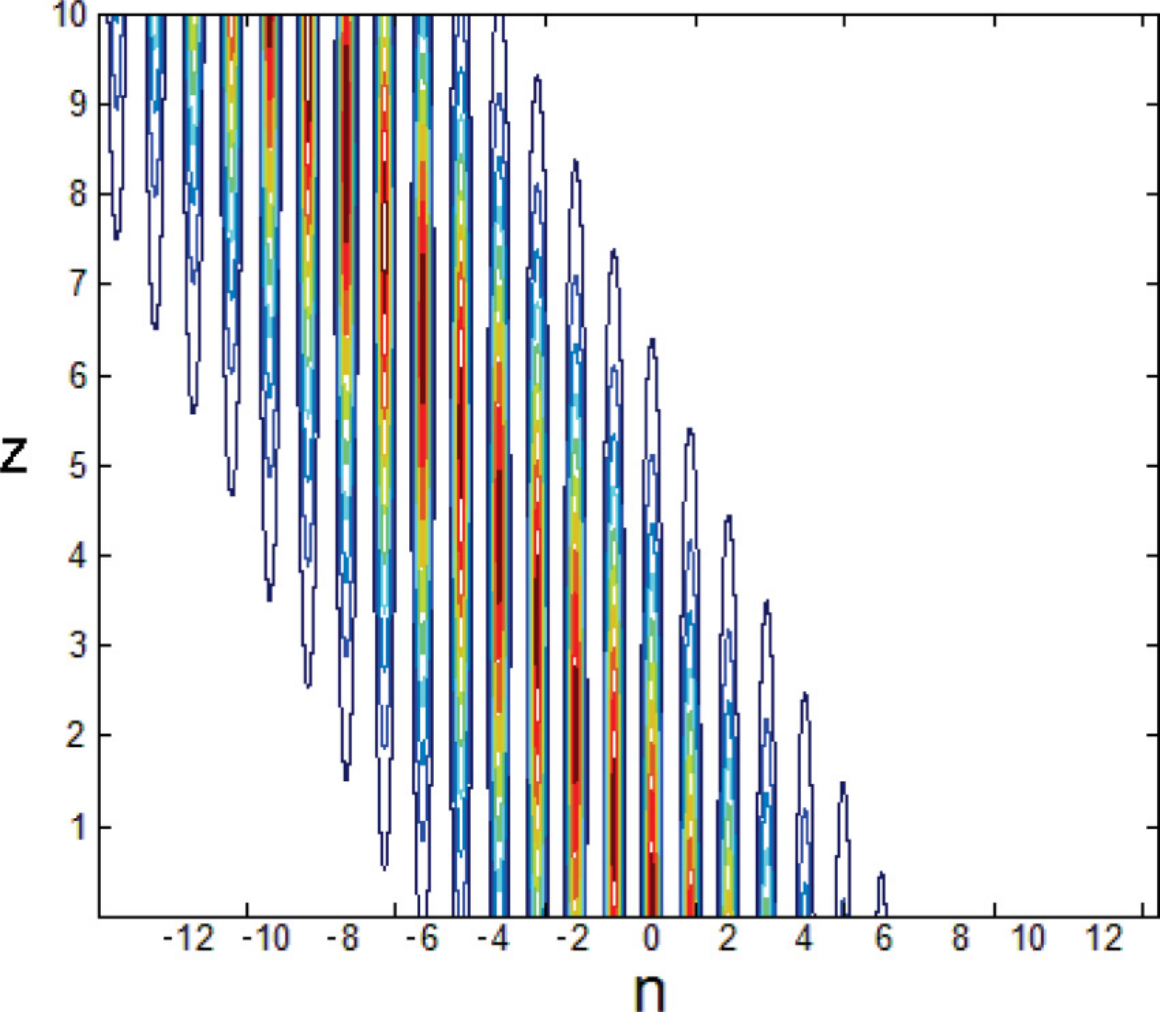}
\caption{An example of a semi-discrete localized spatiotemporal mode,
generated by Eq. (\protect\ref{kappa}), which performs stable transverse
motion under the action of the kick, defined according to (\protect\ref{kick}%
), with $a=1.5$. The cross section of the plot at any fixed $z$ shows the
distribution of power $\left\vert u_{n}(\protect\tau )\right\vert ^{2}$ for
each $n$. The figure is borrowed from Ref. \protect\cite{Blit}.}
\label{fig11}
\end{figure}

A remarkable counter-intuitive property of semi-discrete localized modes
generated by Eq. (\ref{kappa}) is their ability to stably move \emph{across
the array}, under the action of a kick applied at $z=0$ \cite{Blit}:%
\begin{equation}
u_{n}(\tau )\rightarrow \exp \left( ian\right) u_{n}(\tau ),  \label{kick}
\end{equation}%
with real $a$, in spite of the presence of the respective quasi-1D
Peierls-Nabarro potential, An example of such a moving mode is displayed in
Fig. \ref{fig11}. This property may be compared to the above-mentioned
mobility of 1D discrete solitons in the DNLS equation \cite{Feddersen}, and
of 2D discrete solitons in the framework of the $\chi ^{(2)}$ system (\ref%
{geq4}).

\begin{figure}
\includegraphics[width=12cm,scale=1]{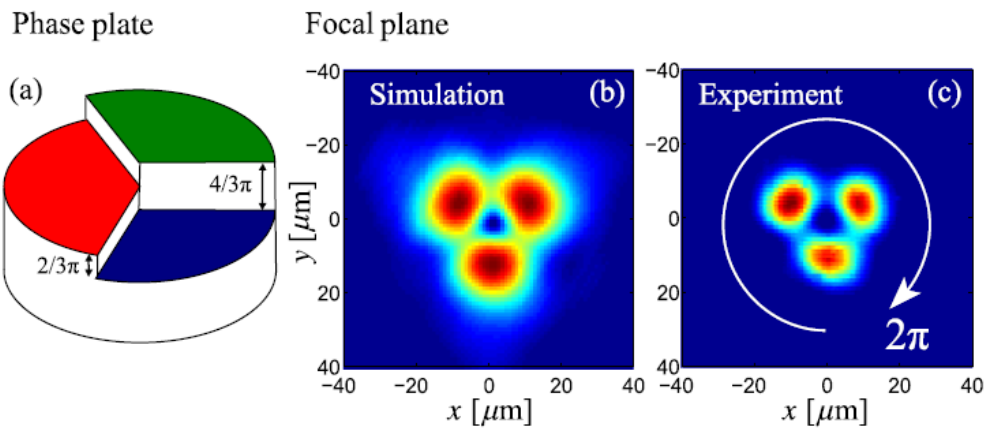} 
\caption{A semi-discrete vortex soliton in a hexagonal quasi-discrete array
of waveguides made in bulk silica. (a) The phase plate used for imprinting
the vortex structure into the input beam. (b,c) Numerically simulated and
experimentally observed (transient) intensity distributions in the
transverse plane, with phase shifts $2\protect\pi /3$ between adjacent
peaks, which represents the simplest vortical phase structure with
topological charge $S=1$. The figure is borrowed from Ref. \protect\cite%
{Jena-vort}.}
\label{fig12}
\end{figure}

Similarly, quasi-discrete settings modeled by an extension of (\ref{kappa})
with two transverse spatial coordinates were used for the creation for
spatiotemporal optical solitons (\textquotedblleft light bullets") \cite%
{Jena}, as well as soliton-like transient modes with embedded vorticity \cite%
{Jena-vort}. Waveguides employed in those experiments feature a transverse
hexagonal-lattice structure, written in bulk silica by means of an optical
technology. A spatiotemporal\ vortex state observed in the bundle-like
structure (in the experiment, it is actually a transient one) is represented
by Fig. \ref{fig12}, which displays both numerically predicted and
experimentally observed distributions of intensity of light in the
transverse plane, together with a phase plate used in the experiment to
embed the vorticity into the incident spatiotemporal pulse which was used to
create the mode.

\subsection{Semi-discrete quantum and photonic droplets}

A new type of semi-discrete solitons was recently elaborated in Ref. \cite%
{Raymond}, in the framework of an array of linearly coupled 1D GP equations,
including the above-mentioned Lee-Huang-Yang correction, which represents an
effect of quantum fluctuations around the mean-field states of a binary BEC
\cite{Petrov,Petrov-Astra}. The system is%
\begin{equation}
i\partial _{t}\psi _{j}=-{(1/2)}\partial _{zz}\psi _{j}-\left( C/2\right)
\left( \psi _{j+1}-2\psi _{j}+\psi _{j-1}\right) +g|\psi _{j}|^{2}\psi
_{j}-|\psi _{j}|\psi _{j},  \label{GPE array}
\end{equation}%
where $\psi _{j}(z)$ is the mean-field wave function in the $j$-th core with
coordinate $z$, $C$ is the effective inter-core coupling constant, the
self-attractive quadratic term represents the Lee-Huang-Yang correction in
the 1D limit, cf. Eq. (\ref{amended 1D}), and $g>0$ accounts for the
mean-field self-repulsion.

\begin{figure}
\includegraphics[width=12cm,scale=1]{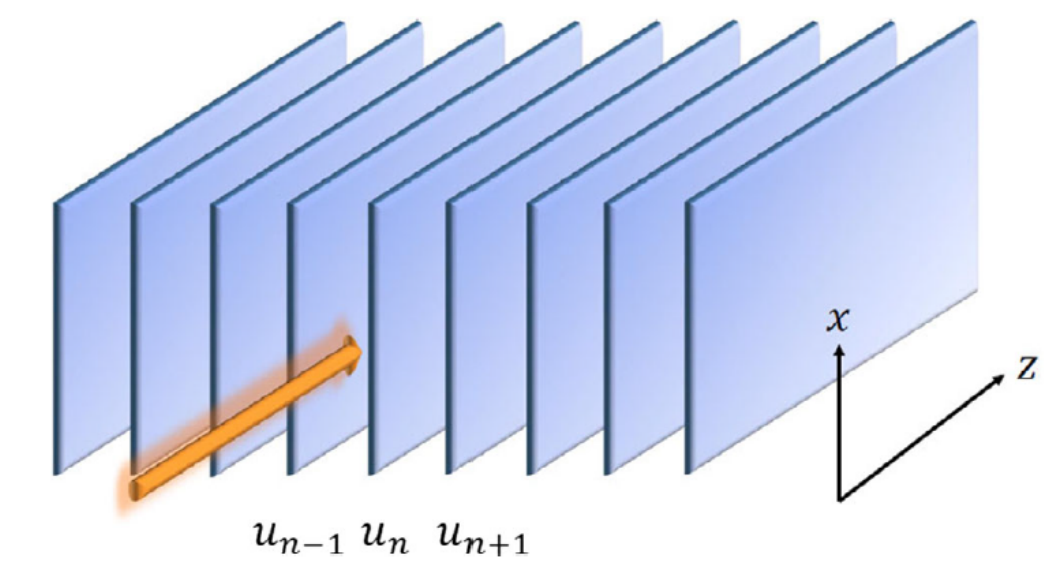}
\caption{The realization of the semi-discrete system \protect\ref{opt array}%
: the array of planar optical waveguides (blue slabs), separated by gray
isolating layers, with the continuous transverse coordinate, $x$, and the
discrete one, $n$. As shown by the arrow, light is coupled into the array
along the $z$ direction. The figure is borrowed from Ref. \protect\cite%
{Raymond2}.}
\label{fig19}
\end{figure}

A semi-discrete system similar to the one modeled by Eq. (\ref{GPE array}),
but with the cubic-quintic nonlinearity instead of the combination of the
quadratic and cubic terms in Eq. (\ref{GPE array}), was derived in the
context of nonlinear optics \cite{Raymond2}:%
\begin{equation}
i\partial _{z}u_{n}=-{(1/2)}\partial _{xx}u_{n}-{(C/2)}\left(
u_{n+1}-2u_{n}+u_{n-1}\right) -|u_{n}|^{2}u_{n}+|u_{n}|^{4}u_{n}.
\label{opt array}
\end{equation}%
It corresponds to the array of parallel-coupled planar waveguides, as shown
in Fig. \ref{fig19}. In this case, $u_{n}(x,z)$ is the complex local
amplitude of the optical wave in the $n$-th waveguide, $z$ is the
propagation distance, and $x$ is the transverse coordinate in each
waveguide, while $C$ is the effective coupling constant, similar to Eq. (\ref%
{GPE array}). By analogy with the quantum droplets, semi-discrete solitons
produced by Eq. (\ref{opt array}) may be called \textquotedblleft photonic
droplets".

The droplets produced by Eqs. (\ref{GPE array}) and (\ref{opt array}) are
characterized by the total norm, which is proportional to the number of
atoms in BEC,%
\begin{equation}
N=\sum_{j}\int_{-\infty }^{+\infty }\left\vert \psi _{j}(z)\right\vert
^{2}dz,  \label{Norm}
\end{equation}%
or the total power of the photonic droplet,%
\begin{equation}
P=\sum_{n}\int_{-\infty }^{+\infty }\left\vert u_{n}(x)\right\vert ^{2}dx.
\label{Power}
\end{equation}%
For solitons produced by Eqs. (\ref{GPE array}) and (\ref{opt array}) sets
of control parameters are, respectively, $\left( C,g\right) $ for fixed $N$,
or $\left( C,P\right) $.

\begin{figure}
\begin{center}
\includegraphics[width=12cm,scale=1]{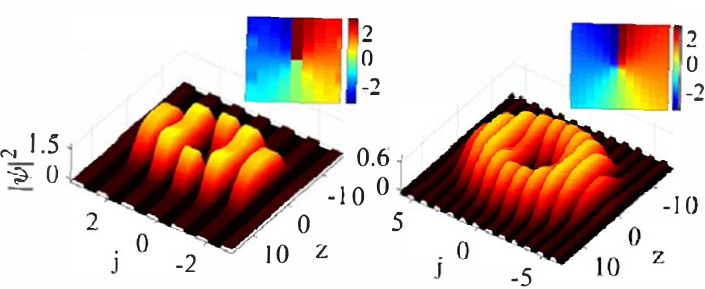}
\end{center}
\caption{Left and right panels display amplitude and phase profiles of
stable onsite- and intersite-centered semi-discrete vortex solitons with
topological charge $S=1$, produced by the numerical solution of Eq. (\protect
\ref{GPE array}) with parameters $\left( g,C\right) =\left( 0.48,0.1\right) $
and $\left( 0.77,0.15\right) $, respectively. The norm (\protect\ref{Norm})
of both solutions is $N=100$. The figure is borrowed from Ref. \protect\cite%
{Raymond}.}
\label{fig13}
\end{figure}

\begin{figure}
\begin{center}
\includegraphics[width=10cm,scale=1]{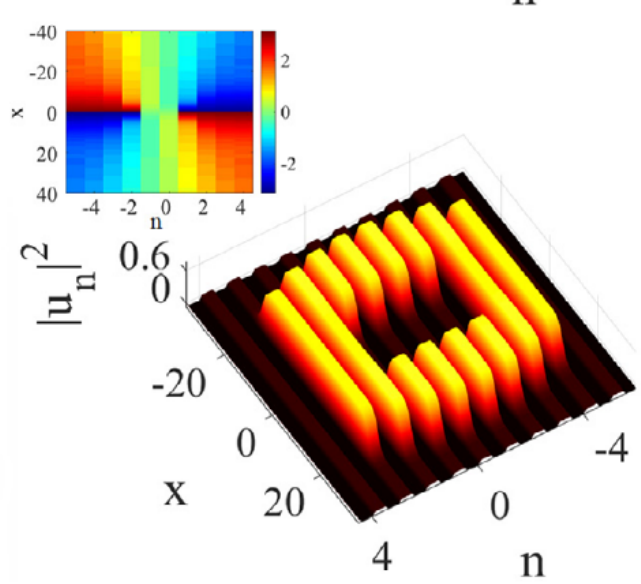}
\end{center}
\caption{Amplitude and phase profiles of a stable intersite-centered
semi-discrete vortex soliton with topological charge $S=2$, produced by the
numerical solution of Eq. (\protect\ref{opt array}) with $C=0.018$. The
total power (\protect\ref{Power}) of this soliton is $P=250$. The figure is
borrowed from Ref. \protect\cite{Raymond2}.}
\label{fig21}
\end{figure}

The models based on Eqs. (\ref{GPE array}) and (\ref{opt array}) give rise
to many families of semi-discrete solitons, including a novel species of
semi-discrete vortex solitons. Typical examples of the onsite- and intersite
vortices with topological charge $S=1$, produced by Eq. (\ref{GPE array}),
are displayed in Fig. \ref{fig13}.
An example of a stable semi-discrete vortex soliton produced by Eq. (\ref%
{opt array}) with $S=2$ is displayed in Fig. \ref{fig21}.

\begin{figure}
\begin{center}
\includegraphics[width=10cm,scale=1]{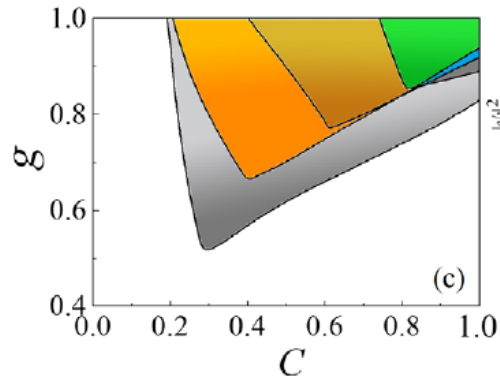}
\end{center}
\caption{Stability areas in the parameter plane $\left( C,g\right) $,
produced by the numerical solution of Eq. (\protect\ref{GPE array}) for
onsite-centered semi-discrete vortex solitons with $S=2$ (all colored
regions), $3$ (orange $+$ brown $+$ green), $4$ (brown $+$ green $+$ blue),
and $5$ (green $+$ blue $+$ dark gray). For the convenience of plotting, the
normalizations for $S=2$, $3$, $4$, and $5$ are fixed as $N=400$, $900$, $%
2500$, and $4500$, respectively. The figure is borrowed from Ref.
\protect\cite{Raymond}.}
\label{fig20}
\end{figure}

Getting back to the semi-discrete system (\ref{GPE array}), a chart in the
plane of $\left( C,g\right) $ which displays stability areas for the
semi-discrete vortex solitons with topological charges $S=2$, $3$, $4$, and $%
5$, is plotted in Fig. \ref{fig20}. The chart demonstrates abundant
multistability \ -- for instance, the stable solitons with $S=2$ \ coexist
with all higher-order ones (with $S=3$, $4$, and $5$).

Self-trapped solutions of a continuum model, which are similar to
semi-discrete vortex solitons outlined above, were recently reported for a
photonic crystal built in a $\chi ^{(2)}$ material with a checkerboard
structure representing quasi-phase matching \cite{Raymond3}.

Semi-discreteness of another type is possible in two-component systems,
where one component is governed by a discrete equation, and the other one by
a continuous equation. This type of two-component systems was proposed in
\cite{Panoiu}. It introduced a $\chi ^{(2)}$ model, assuming that the
continuous second-harmonic wave propagates in a slab with a continuous
transverse coordinate, while the fundamental-harmonic field is trapped in a
discrete waveguiding array built on top of the slab.

\section{2D fundamental and vortical discrete solitons in a two-component $%
\mathcal{PT}$ (parity-time) symmetric lattice}

While the above presentation deals solely with conservative discrete
systems, many properties of conservative settings are shared by a very
special type of dissipative ones, \textit{viz}., systems with the
parity-time ($\mathcal{PT}$) symmetry. They include mutually symmetric
spatially separated elements carrying linear gain and loss \cite%
{Bender,Christod,Christod2}. The experimental realization of such systems in
optics \cite{Christod2} suggests one to include the Kerr nonlinearity, thus
opening the way to the prediction an creation of $\mathcal{PT}$-symmetric
solitons \cite{PTrev1,PTrev2}. In particular, exact solutions for 1D $%
\mathcal{PT}$-symmetric solitons and exact results for their stability
boundaries were found in the model of the nonlinear $\mathcal{PT}$-symmetric
coupler (dual-core waveguide), with mutually symmetric linear gain and loss
carried by the linearly coupled cores \cite{Radik,Barash}. Stability limits
for 2D fundamental solitons in the 2D $\mathcal{PT}$-symmetric coupler with
the cubic-quintic nonlinearity in each core (essentially the same as in Eqs.
(\ref{opt array}), chosen to prevent the critical-collapse instability)
were identified in Ref. \cite{Gena}.

The definition of the $\mathcal{PT}$ symmetry makes it also natural to
consider discrete $\mathcal{PT}$-symmetric systems. Various species of
stable discrete solitons were predicted in chains of $\mathcal{PT}$%
-symmetric elements \cite{PTsol1,PTsol0,PTsol2,PT,PTsol4,PTsol5}, and the
existence of such solitons was demonstrated experimentally \cite%
{PTsol-observation}.

A natural model for the creation of $\mathcal{PT}$-symmetric discrete 2D
solitons is a generalization of the 2D discrete nonlinear coupler, based on
Eqs. (\ref{coupled}), by adding the linear gain and loss terms with strength
$\gamma >0$ to the coupled equations \cite{PT}:
\begin{eqnarray}
&&i{\frac{d\psi _{m,n}}{dz}}=-{\frac{C}{2}}(\psi _{m,n+1}+\psi _{m,n-1}+\psi
_{m-1,n}+\psi _{m+1,n}-4\psi _{m,n})  \notag \\
&&-|\psi _{m,n}|^{2}\psi _{m,n}-\varphi _{m,n}+i\gamma \psi _{m,n},  \notag
\\
&&i{\frac{d\varphi _{m,n}}{dz}}=-{\frac{C}{2}}(\varphi _{m,n+1}+\varphi
_{m,n-1}+\varphi _{m-1,n}+\varphi _{m+1,n}-4\varphi _{m,n})  \notag \\
&&-|\varphi _{m,n}|^{2}\varphi _{m,n}-\psi _{m,n}-i\gamma \varphi _{m,n}.
\label{basicEq}
\end{eqnarray}%
Here, in terms of the optical realization, the evolution variable $z$ is the
propagation distance, the inter-core coupling coefficient is scaled to be $1$%
, and $C>0$ is constant of the intra-core coupling between adjacent sites of
the lattice. The dispersion relations for discrete plane-wave solutions to the
linearized version of Eqs. (\ref{basicEq}), $\left\{ \psi _{m,n}(z),\varphi
_{m,n}(z)\right\} \sim \exp \left( ipm+iqn+ikz\right) $, is%
\begin{equation}
k=-2C\left( \sin ^{2}\frac{p}{2}+\sin ^{2}\frac{q}{2}\right) \pm \sqrt{%
1-\gamma ^{2}}.  \label{k}
\end{equation}%
As it follows from Eq. (\ref{k}), the $\mathcal{PT}$ symmetry holds under
condition $\gamma <\gamma _{\max }\equiv 1$, i.e., the gain-loss strength $%
\gamma $ must be smaller than the linear-coupling coefficient, that is $1$
in the present notation, which is a generic property of $\mathcal{PT}$%
-symmetric couplers \cite{Radik,Barash}.

Stationary modes with real propagation constant $k$ are looked for as
solutions to the full nonlinear system of Eqs. (\ref{basicEq}) in the usual
form, $\left\{ \psi _{m,n}(z),\varphi _{m,n}(z)\right\} =e^{ikz}\left\{
u_{m,n},v_{m,n}\right\} $, with stationary amplitudes obeying equations%
\begin{eqnarray}
\left( k+i\gamma \right) u_{m,n} &=&{\frac{C}{2}}%
(u_{m,n+1}+u_{m,n-1}+u_{m-1,n}+u_{m+1,n}-4u_{m,n})+|u_{m,n}|^{2}u_{m,n}+v_{m,n},
\notag \\
\left( k-i\gamma \right) v_{m,n} &=&{\frac{C}{2}}%
(v_{m,n+1}+v_{m,n-1}+v_{m-1,n}+v_{m+1,n}-4v_{m,n})+|v_{m,n}|^{2}v_{m,n}+u_{m,n}.
\label{UV}
\end{eqnarray}%
Localized states produced by Eqs. (\ref{UV}) are characterized, as above, by
the total power,
\begin{equation}
P=\sum {}_{m,n}\left( |u_{m,n}|^{2}+|v_{m,n}|^{2}\right) .  \label{P}
\end{equation}

\begin{figure}
{\includegraphics[scale=0.8]{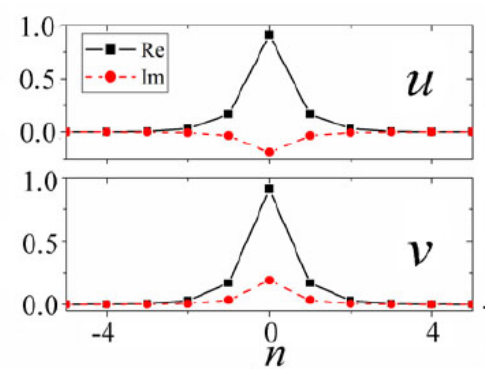}}
\caption{Cross-sections, along $m=0$, of real and imaginary parts of
components $u_{m,n}$ and $v_{m,n}$ of a stable $\mathcal{PT}$-symmetric
fundamental soliton produced by the numerical solution of Eq. (\protect\ref%
{w}), with $(C,\protect\gamma )=(0.3,0.4)$, taking relation (\protect\ref%
{PTsymm}) into regard. The total power (\protect\ref{P}) of the soliton is $%
P=2$. The figure is borrowed from Ref. \protect\cite{PT}.}
\label{fig22}
\end{figure}

Straightforward analysis of Eqs. (\ref{UV}) demonstrates that the system
produces $\mathcal{PT}$-symmetric fundamental-soliton solutions, which must
be subject to the relation $v_{m,n}=u_{m,n}^{\ast }$ (with $\ast $ standing
for the complex conjugate), in the form of%
\begin{equation}
\left\{ u_{m,n},v_{mn}\right\} =w_{m,n}\exp \left( \pm (i/2)\arcsin \gamma
\right) ,  \label{PTsymm}
\end{equation}%
where real discrete distribution $w_{m,n}$ should be found as a solution of
the usual stationary equation for 2D discrete solitons,%
\begin{equation}
\left( -k+\sqrt{1-\gamma ^{2}}\right) w_{m,n}+{\frac{C}{2}}\left(
w_{m,n+1}+w_{m,n-1}+w_{m-1,n}+w_{m+1,n}-4w_{m,n}\right) +w_{m,n}^{3}=0,
\label{w}
\end{equation}%
cf. Eq. (\ref{2D-DNLSE}). In agreement with the linear spectrum (\ref{k}),
Eq. (\ref{w}) may produce soliton solutions for $k>\sqrt{1-\gamma ^{2}}$. An
example of a stable fundamental $\mathcal{PT}$-symmetric soliton is
displayed, by means of its cross-section shapes, in Fig. \ref{fig22}.

\begin{figure}
{\includegraphics[scale=0.8]{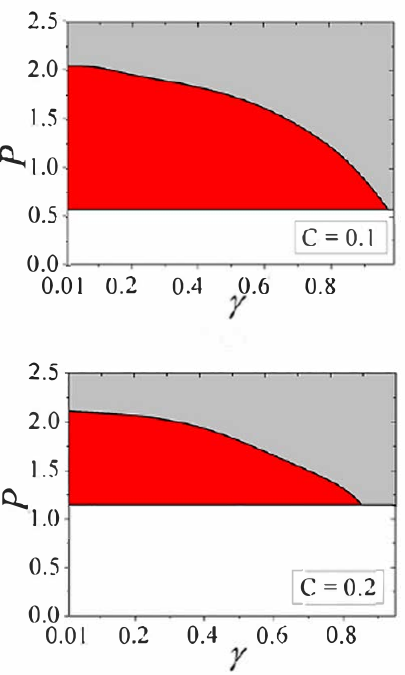}}
\caption{Red and gray colors designate, respectively, stability and
instability areas for $\mathcal{PT}$-symmetric fundamental discrete
solitons, produced by Eqs. (\protect\ref{w}) and (\protect\ref{PTsymm}), in
the plane of the gain-loss coefficient, $\protect\gamma $, and total power $%
P $, which is defined as per Eq. (\protect\ref{P}). The soliton solutions do
not exist in the white area. The figure is borrowed from Ref. \protect\cite%
{PT}.}
\label{fig23}
\end{figure}

The existence and stability of the $\mathcal{PT}$-symmetric fundamental
discrete solitons is summarized in the plane of $\left( \gamma ,P\right) $
for the fundamental solitons in Fig. \ref{fig23}. It \ is seen that,
naturally, the stability area shrinks as the gain-loss coefficient $\gamma $
is approaching its limit value, $\gamma _{\max }=1$ (cf. the 1D situation
considered in Refs. \cite{Radik} and \cite{Barash}). The existence boundary,
i.e., the minimum value of $P$, below which no solitons are found (in the
white area), corresponds to the limit of very broad small-amplitude
solitons. In this limit, the discrete soliton may be approximated by its
counterpart in the continuum NLS\ equation, i.e., the above-mentioned Townes
soliton, whose power takes the unique value, which thus determines the
existence boundary in Fig. \ref{fig23}.

The stability boundary in Fig. \ref{fig23} may be understood as the one at
which the symmetric soliton is destabilized by the spontaneous symmetry
breaking (as described in detail above for 2D solitons produced by the
linearly-coupled conservative DNLS equations (\ref{coupled}), see also Ref.
\cite{Herring}), which is here modified by the presence of the linear gain
and loss. Because asymmetric solitons cannot exist in the system with the
balanced gain and loss, the symmetry breaking always leads to either blowup
or decay of the soliton \cite{PT}. In their stability region, the $\mathcal{%
PT}$-symmetric fundamental discrete solitons actually represent the system's
ground state \cite{PT}.

\begin{figure}
\centering
\includegraphics[scale=0.7]{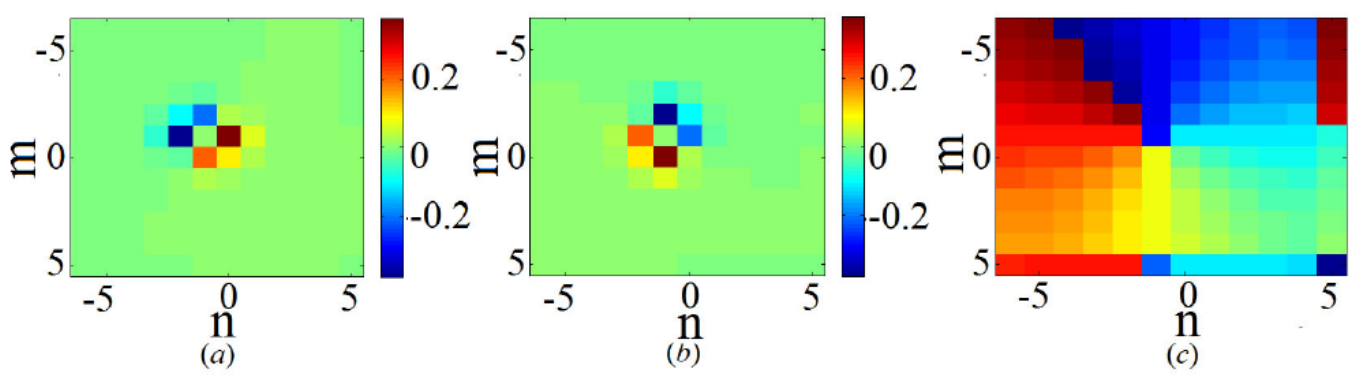}
\caption{Real (a) and imaginary (b) parts, and the phase structure (c), of
field $u_{m,n}$ of a stable discrete $\mathcal{PT}$-symmetric vortex
soliton, produced by Eqs. (\protect\ref{UV}) for $(C,\protect\gamma %
)=(0.06,0.4)$, with propagation constant $k=1$ and total power $P=1.65$,
defined as per Eq. (\protect\ref{P}). The figure is borrowed from Ref.
\protect\cite{PT}.}
\label{fig24}
\end{figure}

Alongside the fundamental discrete $\mathcal{PT}$-symmetric solitons, the
same system of Eqs. (\ref{UV}) produces $\mathcal{PT}$-symmetric vortex
solitons, which also have their stability area, see details in Ref. \cite{PT}%
. An example of a stable $\mathcal{PT}$-symmetric vortex soliton is
presented in \ref{fig24}.

In addition to the $\mathcal{PT}$-symmetric solitons, Eqs. (\ref{UV}) give
rise to anti-$\mathcal{PT}$-symmetric ones, defined by relation $%
v_{m,n}=-u_{m,n}^{\ast }$. They, as well as anti-$\mathcal{PT}$-symmetric
vortex solitons, are stable in some parameter areas (see details in Ref.
\cite{PT}), but those areas are essentially smaller than their counterparts
for the $\mathcal{PT}$-symmetric modes. The reduced stability area for the
anti-$\mathcal{PT}$-symmetric fundamental solitons is explained by the fact
that they cannot be the system's ground state.

\section{Conclusion}

The interplay of the discreteness and intrinsic nonlinearity in various
physical media -- chiefly, in nonlinear optics and BEC -- gives rise to a
great variety of self-trapped localized states, in the form of discrete
solitons. This article aims to produce a concise review, starting from the
brief survey of basic theoretical models combining the discreteness in 1D,
2D, and 3D geometries and various nonlinearities, such as cubic, quadratic,
and quintic. The main subject addressed in the article is a summary of basic
results for 2D and 3D discrete solitons produced by such models. Unlike the
topic of 1D discrete solitons, the multidimensional ones were not previously
reviewed in a systematic form. Along with the fundamental solitons,
topologically organized ones, in the form of solitary vortices and discrete
skyrmions, are considered too. Some experimental findings are also included,
such as the observation of 2D discrete optical solitons with embedded
vorticity.

In many cases, the discreteness helps to produce states which either do not
exist or are definitely unstable in continuum analogs of the discrete
settings. In particular, these are 2D fundamental and vortex solitons, which
may be stable in the discrete form, while their continuum counterparts are
completely unstable in the free space. On the other hand, mobility of
solitons, which is their obvious property in the continuum, is a nontrivial
issue for the lattice (discrete) solitons.

The work in this area remains a subject of ongoing theoretical and
experimental work, promising new findings. A perspective direction is to
produce 2D and 3D self-trapped states with intrinsic topological structures.
Some results obtained in this direction are presented in this article, such
as discrete solitons in the system with spin-orbit coupling \cite{HS} (see
also Ref. \cite{Sandra}), sophisticated 3D discrete modes with embedded
vorticity \cite{3Dsol1,3Dsol2}, and discrete skyrmions \cite{skyrmion}. A
challenging task is experimental realization of these states which, thus
far, were only predicted in the theoretical form.

It is relevant to mention some topics which may be relevant in the present
context but are not included here, to keep a reasonable size of the review.
In particular, these are interactions of discrete solitons with local
defects in the underlying lattice, as well as with interfaces and edges. It
is known that defects and surfaces may often help to create and stabilize
localized modes which do not exist or are unstable in uniform lattices, such
as Tamm \cite{Tamm} and topological-insulator \cite{top-ins,top-ins2}
states. Another vast area of studies, which is not considered here, deals
with dissipative discrete nonlinear systems. In this article, only the very
special case of $\mathcal{PT}$-symmetric systems is addressed. Basic
nonlinear dissipative models are represented by discrete complex
Ginzburg-Landau equations, i.e., DNLS equations with complex coefficients in
front of the onsite linear and nonlinear terms, which account for losses and
compensating gain \cite{Hakim}. Unlike conservative and $\mathcal{PT}$%
-symmetric models, the dissipative ones may only give rise to stable
discrete solitons which do not exist in continuous families, but rather as
isolated \textit{attractors} \cite{Efremidis1,Akhmed,Efremidis2}.

\section*{Acknowledgments}

I appreciate the invitation of Editors of the special issue of Entropy,
Profs. Lars English and Faustino Palmero, to submit this article as a
contribution to the special issue on the topic of \textquotedblleft Recent
Advances in the Theory of Nonlinear Lattices". I would like to thank
colleagues in collaboration with whom I have been working on various topics
related to the review: G. E. Astrakharchik, B. B. Baizakov, P. Beli\v{c}ev,
A. R. Bishop, R. Blit, L. L. Bonilla, O. V. Borovkova, R. Carretero-Gonz\'{a}%
lez, Zhaopin Chen, Zhigang Chen, C. Chong, J. Cuevas-Maraver, J. D'Ambroise,
F. K. Diakonos, S. V. Dmitriev, R. Driben, N. Dror, O. Dutta, A. Eckhard, L.
M. Flor\'{\i}a, D. J. Frantzeskakis, S. Fu, G. Gligori\'{c}, J. G\'{o}%
mez-Garde\~{n}es, L. Had\v{z}ievsli, D. Herring, J. Hietarinta, P. Hauke, T.
Kapitula, Y. V. Kartashov, D. J. Kaup (deceased). P. G. Kevrekidis, V. V.
Konotop, M. Lewenstein, Ben Li, Yongyao Li, A. Maluckov, T. Meier, D.
Mihalache, N. C. Panoiu, I. E. Papachalarampus, J. Petrovi\'{c}, M. A.
Porter, K. \O . Rasmussen, H. Sakaguchi, M. Salerno, B. S\'{a}nchez-Rey, H.
Susanto, L. Torner, M. Trippenbach, A. V. Ustinov, R. A. Van Gorder, A.
Vardi, M. I. Weinstein, X. Xu, A. V. Yulin, and K. Zegadlo.

This work was supported, in part, by the Israel Science Foundation, through
grant No. 1695/22.


\begin{thebibliography}{999}
\bibitem{Pit} L. P. Pitaevskii and S. Stringari, \textit{Bose-Einstein
Condensation} (Oxford University Press, Oxford, 2003).

\bibitem{KA} Yu. S. Kivshar, G. P. Agrawal, \textit{Optical Solitons: From
Fibers to Photonic Crystals} (Academic Press, San Diego, 2003).

\bibitem{Gadi} G. Fibich, \textit{The Nonlinear Schr\"{o}dinger Equation:
Singular Solutions and Optical Collapse} (Springer, Heidelberg, 2015).

\bibitem{OL} O. Morsch and M. Oberthaler, Dynamics of Bose-Einstein
condensates in optical lattice, Rev. Mod. Phys. \textbf{78}, 179-215 (2006).

\bibitem{Porter} M. A. Porter, R. Carretero-Gonz\'{a}lez, P. G. Kevrekidis,
and B.A. Malomed, Nonlinear lattice dynamics of Bose-Einstein condensates,
Chaos \textbf{15}, 015115 (2005).

\bibitem{PhotCryst} M. Skorobogatiy, J. Yang, \textit{Fundamentals of
Photonic Crystal Guiding} (Cambridge University Press, Cambridge, 2008)

\bibitem{ChristoJoseph} D. N. Christodoulides and R. I. Joseph, Discrete
self-focusing in nonlinear arrays of coupled waveguides, Opt. Lett. \textbf{%
13}, 794-796 (1988)

\bibitem{Silberberg} H. S. Eisenberg, Y. Silberberg, R. Morandotti, A. R.
Boyd, and J. S. Aitchison, Discrete spatial optical solitons in waveguide
arrays, Phys. Rev. Lett. \textbf{81}, 3383-3386 (1998).

\bibitem{discr-review-0} D. N. Christodoulides, F. Lederer, and Y.
Silberberg, Discretizing light behaviour in linear and nonlinear waveguide
lattices, Nature \textbf{424}, 817-823 (2003).

\bibitem{discr-review} F. Lederer, G. I. Stegeman, D. N. Christodoulides, G.
Assanto, M. Segev, and Y. Silberberg, Discrete solitons in optics, Phys.
Rep. \textbf{463}, 1-126 (2008).

\bibitem{Nicolae} F. Ye, D. Mihalache, B. Hu, N. C. Panoiu, Subwavelength
plasmonic lattice solitons in arrays of metallic nanowires, Phys. Rev. Lett.
\textbf{104}, 106802 (2010).

\bibitem{Smerzi} A. Smerzi and A. Trombettoni, Nonlinear tight-binding
approximation for Bose-Einstein condensates in a lattice, Phys. Rev. A
\textbf{68}, 023613 (2003).

\bibitem{Wannier} G. Alfimov, P. Kevrekidis, V. Konotop, and M. Salerno,
Wannier functions analysis of the nonlinear Schr\"{o}dinger equation with a
periodic potential, Phys. Rev. E \textbf{66} 046608 (2002).

\bibitem{Wannier-review} N. Marzari, A. A. Mostofi, J. R. Yates, I, Souza,
and D. Vanderbilt, Maximally localized Wannier functions: Theory and
applications, Rev. Mod. Phys. \textbf{84}, 1419-1475 (2012).

\bibitem{NNN} A. Szameit, R. Keil, F. Dreisow, M. Heinrich, T. Pertsch, S.
Nolte, and A. T\"{u}nnermann, Observation of discrete solitons in lattices
with second-order interaction, Opt. Lett. \textbf{34}, 2838-2840 (2009).

\bibitem{ChongNNN} C. Chong, R. Carretero-Gonz\'{a}lez, B. A. Malomed, and
P. G. Kevrekidis, Variational approximations in discrete nonlinear Schr\"{o}%
dinger equations with next-nearest-neighbor couplings, Physica D \textbf{240}%
, 1205-1212 (2011).

\bibitem{NNN2D} A. Szameit, T. Pertsch, S. Nolte, A. T\"{u}nnermann, F.
Lederer, Long-range interaction in waveguide lattices, Phys. Rev. A \textbf{%
77}, 043804 (2008).

\bibitem{Angelis} A. Locatelli, D. Modotto, D. Paloschi, and C. De Angelis,
All optical switching in ultrashort photonic crystal couplers, Opt. Commun.
\textbf{237}, 97-102 (2004).

\bibitem{Herring} G. Herring, P. G. Kevrekidis, B. A. Malomed, R.
Carretero-Gonz\'{a}lez, D. J. Frantzeskakis, Symmetry breaking in linearly
coupled dynamical lattices, Phys. Rev. E \textbf{76}, 066606 (2007).

\bibitem{Aubry} S. Aubry, Breathers in nonlinear lattices: Existence, linear
stability and quantization, Physica D \textbf{103}, 201-250 (1997).

\bibitem{Kartashov} Y. V. Kartashov, V. A. Vysloukh, and L. Torner, Soliton
shape and mobility control in optical lattices, Prog. in Optics \textbf{52},
63-148 (2009).

\bibitem{Rothos} V. M. Rothos, Nonlinear wave propagation in discrete and
continuous systems, Eur. Phys. J. Special Topics \textbf{225}, 943--958
(2016).

\bibitem{Tsoy} E. N. Tsoy and B. A. Umarov, Introduction to nonlinear
discrete systems: theory and modelling, Eur. J. Phys. \textbf{39}, 055803
(2018).

\bibitem{chapter} B. A. Malomed, Nonlinearity and discreteness: Solitons in
lattices, in \textit{Emerging Frontiers in Nonlinear Science}, pp. 81-110,
ed. by P. G. Kevrekidis, J. Cuevas-Maraver, and A. Saxena (Springer Nature
Switzerland AG: Cham, 2020).

\bibitem{DNLS-book} P. G. Kevrekidis, \textit{The Discrete Nonlinear Schr%
\"{o}dinger Equation: Mathematical Analysis, Numerical Computations, and
Physical Perspectives} (Springer, Berlin Heidelberg, 2009).

\bibitem{Laedke} E. W. Laedke, K. H. Spatschek, and S. K. Turitsyn,
Stability of discrete solitons and quasicollapse to intrinsically localized
modes, Phys. Rev. Lett. \textbf{73}, 1055-1059 (1994).

\bibitem{NRP} Y. Kartashov, G. Astrakharchik, B. Malomed, and L. Torner,
Frontiers in multidimensional self-trapping of nonlinear fields and matter,
Nature Reviews Physics \textbf{1}, 185-197 (2019).

\bibitem{book} B. A. Malomed, \textit{Multidimensional Solitons} (AIP
Publishing, Melville, NY, 2022).

\bibitem{X} B. A. Malomed, Multidimensional Soliton Systems, Advances in
Physics: X, in press.

\bibitem{LHY} T. D. Lee, K. Huang, and C. N. Yang, Eigenvalues and
eigenfunctions of a Bose system of hard spheres and its low temperature
properties, Phys. Rev. \textbf{106}, 1135-1145 (1957).

\bibitem{Petrov} D. S. Petrov, Quantum mechanical stabilization of a
collapsing Bose-Bose mixture, Phys. Rev. Lett. \textbf{115}, 155302 (2015).

\bibitem{Tarruell1} C. Cabrera, L. Tanzi, J. Sanz, B. Naylor, P. Thomas, P.
Cheiney, and L. Tarruell, Quantum liquid droplets in a mixture of
Bose-Einstein condensates, Science \textbf{359}, 301-304 (2018).

\bibitem{Tarruell2} P. Cheiney, C. R. Cabrera, J. Sanz, B. Naylor, L. Tanzi,
and L. Tarruell, Bright soliton to quantum droplet transition in a mixture
of Bose-Einstein condensates, Phys. Rev. Lett. \textbf{120}, 135301 (2018).

\bibitem{Inguscio} G. Semeghini, G. Ferioli, L. Masi, C. Mazzinghi, L.
Wolswijk, F. Minardi, M. Modugno, G. Modugno, M. Inguscio, and M. Fattori,
Self-bound quantum droplets of atomic mixtures in free space?, Phys. Rev.
Lett. \textbf{120}, 235301 (2018).

\bibitem{Petrov-Astra} D. S. Petrov and G. E. Astrakharchik, Ultradilute
low-dimensional liquids, Phys. Rev. Lett. \textbf{117}, 100401 (2016).

\bibitem{Grisha} G. E. Astrakharchik and B. A. Malomed, Dynamics of
one-dimensional quantum droplets, Phys. Rev. A \textbf{98}, 01363 (2018).

\bibitem{Zakh} V. E. Zakharov, S. V. Manakov, S. P. Novikov, and L. P.
Pitaevskii, \textit{Solitons: The Inverse Scattering Method} (Nauka
Publishers, Moscow, 1980) [English translation: Consultants Bureau, New
York, 1984]

\bibitem{Segur} M. J. Ablowitz and H. Segur, \textit{Solitons and Inverse
Scattering Method} (SIAM, Philadelphia, 1981)

\bibitem{Calogero} F. Calogero and A. Degasperis, \textit{Spectral Transform
and Solitons: Tools to Solve and Investigate Nonlinear Evolution Equations}
(North-Holland, New York, 1982)

\bibitem{Newell} A. C. Newell, \textit{Solitons in Mathematics and Physics}
(SIAM: Philadelphia, 1985)

\bibitem{Herbst} M. J. Ablowitz and B. M. Herbst, On homoclinic structure
and numerically induced chaos for the nonlinear Schr\"{o}dinger equation,
SIAM J. Appl. Math. \textbf{50}, 339-351 (1990).

\bibitem{non-integrable} D. Levi, M. Petrera, and C. Scimiterna, On the
integrability of the discrete nonlinear Schr\"{o}dinger equation, Europhys.
Lett. \textbf{84}, 10003 (2008).

\bibitem{AL} M. J. Ablowitz and J. F. Ladik, Nonlinear
differential--difference equations and Fourier analysis, J. Math. Phys.
\textbf{17}, 1011-1018 (1976).

\bibitem{Suris} Yu. B. Suris, \textit{The problem of integrable
discretization: Hamiltonian approach} (Birkhauser, Basel, 2003).

\bibitem{SA} M. Salerno, A new method to solve the quantum Ablowitz-Ladik
system, Phys. Lett. A \textbf{162}, 381-384 (1992).

\bibitem{BH-review} O. Dutta, M. Gajda, P. Hauke, M. Lewenstein, D.-S.
Luhmann, B.A. Malomed, T. Sowinski, J. Zakrzewski, Non-standard Hubbard
models in optical lattices: a review, Rep. Prog. Phys. \textbf{78}, 066001
(2015).

\bibitem{Borovkova} O. V. Borovkova, Y. V. Kartashov, L. Torner, and B. A.
Malomed, Bright solitons from defocusing nonlinearities, Phys. Rev. E
\textbf{84}, 035602 (R) (2011).

\bibitem{Gligorich} G. Gligori\'{c}, A. Maluckov, L. Hadzievski, and B. A.
Malomed, Discrete localized modes supported by an inhomogeneous defocusing
nonlinearity, Phys. Rev. E \textbf{88}, 032905 (2013).

\bibitem{center-periphery} P. G. Kevrekidis, B. A. Malomed, A. Saxena, A. R.
Bishop, and D. J. Frantzeskakis, Solitons and vortices in two-dimensional
discrete nonlinear Schr\"{o}dinger systems with spatially modulated
nonlinearity, Phys. Rev. E \textbf{91}, 043201 (2015).

\bibitem{DD-review} T. Lahaye, C. Menotti, L. Santos, M. Lewenstein, and T.
Pfau, The physics of dipolar bosonic quantum gases, Rep. Prog. Phys. \textbf{%
72}, 126401 (2009).

\bibitem{DD-2D} G. Gligori\'{c}, A. Maluckov, M. Stepi\'{c}, L. Had\v{z}%
ievski, and B. A. Malomed, Discrete vortex solitons in dipolar Bose-Einstein
condensates, J. Phys. B: At. Mol. Opt. Phys. \textbf{43}, 055303 (2010).

\bibitem{Pedri} P. Pedri and L. Santos, Two-dimensional bright solitons in
dipolar Bose-Einstein condensates, Phys. Rev. Lett. \textbf{95}, 200404
(2005).

\bibitem{Ami1} I. Tikhonenkov, B. A. Malomed, and A. Vardi, Anisotropic
solitons in dipolar Bose-Einstein condensates,. Phys. Rev. Lett. \textbf{100}%
, 090406 (2008).

\bibitem{QQ} Y. Li, J. Liu, W. Pang, and B. A. Malomed, Lattice solitons
with quadrupolar intersite interactions, Phys. Rev. A \textbf{88}, 063635
(2013).

\bibitem{Buryak} V. Buryak, P. Di Trapani, D. V. Skryabin, and S. Trillo,
Optical solitons due to quadratic nonlinearities: from basic physics to
futuristic applications, Phys. Rep. \textbf{370}, 63-235 (2002).

\bibitem{Hadi} H. Susanto, P. G. Kevrekidis, R. Carretero-Gonz\'{a}lez, B.
A. Malomed, and D. J. Frantzeskakis, Mobility of discrete solitons in
quadratically nonlinear media, Phys. Rev. Lett. \textbf{99}, 214103 (2007).

\bibitem{Zaragoza} J. G\'{o}mez-Garde\~{n}es, B. A. Malomed, L. M. Flor\'{\i}%
a, A. R. Bishop, Solitons in the Salerno model with competing
nonlinearities, Phys. Rev. E \textbf{73}, 036608 (2006).

\bibitem{Aubry2} D. Chen, S. Aubry, G. P. Tsironis, Breather mobility in
discrete $\varphi^{4}$ nonlinear lattices, Phys. Rev. Lett. \textbf{77},
4776-4779 (1996).

\bibitem{Peyrard} Yu. S. Kivshar and M. Peyrard, Modulational instabilities
in discrete lattices, Phys. Rev. A \textbf{46}, 3198-3205 (1992).

\bibitem{Progress} B. A. Malomed, Variational methods in nonlinear fiber
optics and related fields, Prog. Optics \textbf{43}, 71-193 (2002).

\bibitem{Weinstein} B.A. Malomed and M. I. Weinstein, Soliton dynamics in
the discrete nonlinear Schr\"{o}dinger equation, Phys. Lett. A 220, 91-96
(1996).

\bibitem{Papa} I. E. Papacharalampous, P. G. Kevrekidis, B. A. Malomed, D.
J. Frantzeskakis, Soliton collisions in the discrete nonlinear Schr\"{o}%
dinger equation, Phys. Rev. E \textbf{68}, 046604 (2003).

\bibitem{Dave-VA} D. J. Kaup, Variational solutions for the discrete
nonlinear Schr\"{o}dinger equation, Math. Comput. Simulat. \textbf{69},
322-333 (2005).

\bibitem{Gorder} B. A. Malomed, D. J. Kaup, and R. A. Van Gorder,
Unstaggered-staggered solitons in two-component discrete nonlinear Schr\"{o}%
dinger lattices, Phys. Rev. E \textbf{85}, 026604 (2012).

\bibitem{Needs07} J. Cuevas, G. James, P. G. Kevrekidis, B. A. Malomed, B. S%
\'{a}nchez-Rey, Approximation of solitons in the discrete NLS equation, J.
Nonlinear Math. Phys. \textbf{15 }(Suppl. 3), 124-136 (2008).

\bibitem{Kivshar1} Y. S. Kivshar, W. Krolikowski, and O. A. Chubykalo, Dark
solitons in discrete lattices, Phys. Rev. E \textbf{50}, 5020-5032 (1994).

\bibitem{Konotop} G. L. Alfimov, V. V. Konotop, and M. Salerno, Matter
solitons in Bose-Einstein condensates with optical lattices, Europhys. Lett.
\textbf{58}, 7-13 (2002).

\bibitem{Silberberg2} D. Mandelik, R. Morandotti, J. S. Aitchison, and Y.
Silberberg, Gap solitons in waveguide arrays, Phys. Rev. Lett. \textbf{92},
093904 (2004).

\bibitem{Pengfei} Y. Q. Gao, Y. Lv, Z. F. Feng, and P. F. Li, Unidirectional
flow of the discrete dark solitons and excitation of the discrete X-waves in
$\mathcal{PT}$-symmetric optical waveguide arrays, Romanian Reports in
Physics \textbf{74}, 110 (2022).

\bibitem{Kevrekidis} J. Cuevas, G. James, P. G. Kevrekidis, and K. J. H.
Law, Vortex solutions of the discrete Gross-Pitaevskii equation starting
from the anti-continuum limit, Physica D \textbf{238},\ 1422-1431 (2009).

\bibitem{rigorous} C. Chong, D. E. Pelinovsky, and G. Schneider, On the
validity of the variational approximation in discrete nonlinear Schr\"{o}%
dinger equations, Physica D \textbf{241}, 115-124 (2012).

\bibitem{Feddersen} D. B. Duncan, J. C. Eilbeck, H. Feddersen, and A. D.
Wattis, Solitons on lattices, Physica D \textbf{68}, 1-11 (1993).

\bibitem{Toda} M. Toda, Vibration of a chain with a non-linear interaction,
J. Phys. Soc. Jpn. \textbf{22}, 431-436 (1967).

\bibitem{twisted} S. Darmanyan, A. Kobyakov, and F. Lederer, Stability of
strongly localized excitations in discrete media with cubic nonlinearity, J.
Exp. Theor. Phys. \textbf{86}, 682-686 (1998).

\bibitem{bound states} T. Kapitula, P. G. Kevrekidis, and B. A. Malomed,
Stability of multiple pulses in discrete systems, Phys. Rev. E \textbf{63},
036604 (2001).

\bibitem{bound states 2} D. \ E. Pelinovsky, P. G. Kevrekidis, and D. J.
Frantzeskakis, Stability of discrete solitons in nonlinear Schr\"{o}dinger
lattices, Physica D \textbf{212}, 1-19 (2005).

\bibitem{Bishop} P. G. Kevrekidis, B. A. Malomed, and A. R. Bishop, Bound
states of two-dimensional solitons in the discrete nonlinear Schr\"{o}dinger
equation, J. Phys. A: Math. Gen. \textbf{34}, 9615-9629 (2001).

\bibitem{Blit} R. Blit and B. A. Malomed, Propagation and collisions of
semi-discrete solitons in arrayed and stacked waveguides, Phys. Rev. A
\textbf{86}, 043841 (2012).

\bibitem{Driben} R. Driben, V. V. Konotop, B. A. Malomed, T. Meier, and A.
V. Yulin, Nonlinearity-induced localization in a periodically driven
semidiscrete system, Phys. Rev. E \textbf{97}, 062210 (2018).

\bibitem{Raymond} X. Zhang, X. Xu, Y. Zheng, Z. Chen, B. Liu, C. Huang, B.
A. Malomed, and Y. Li, Semidiscrete quantum droplets and vortices, Phys.
Rev. Lett. \textbf{123}, 133901 (2019).

\bibitem{Raymond2} X. Xu, G. Ou, Z. Chen, B. Liu, B. A. Malomed, and Y. Li,
Semidiscrete vortex solitons, Advanced Photonics Research \textbf{2},
2000082 (2021).

\bibitem{Raymond3} X. Xu, F. Zhao, J. Huang, H. Xiang, L. Zhang, Z. Chen, Z.
Nie, B. A Malomed, and Y. Li, Semidiscrete optical vortex droplets in
quasi-phase-matched photonic crystals, Opt. Exp. \textbf{31}, 38343-38354
(2023).

\bibitem{Panoiu} N. C. Panoiu, R. M. Osgood, and B. A. Malomed,
Semi-discrete composite solitons in arrays of quadratically nonlinear
waveguides, Opt. Lett. \textbf{31}, 1097-1099 (2006).

\bibitem{semidiscr vort Adv Phot} X. Xu, G. Ou, Z. Chen, B. Liu, B. A.
Malomed, and Y. Li, Semidiscrete vortex solitons, Advanced Photonics
Research \textbf{2}, 2000082 (2021).

\bibitem{semidicr opt vort drop} X. Xu, F. Zhao, J. Huang, H. Xiang, L.
Zhang, Z. Chen, Z. Nie, B. A Malomed, and Y. Li, Semidiscrete optical vortex
droplets in quasi-phase-matched photonic crystals, Opt. Exp. \textbf{31},
38343-38354 (2023).

\bibitem{Kivshar} D. N. Neshev, T. J. Alexander, E. A. Ostrovskaya, Yu. S.
Kivshar, H. Martin, I. Makasyuk, and Z. G. Chen, Observation of discrete
vortex solitons in optically induced photonic lattices, Phys. Rev. Lett.
\textbf{92}, 123903 (2004),

\bibitem{Segev} J. W. Fleischer, G. Bartal, O. Cohen, O. Manela, M. Segev,
J. Hudock, and D. N. Christodoulides, Observation of vortex-ring
\textquotedblleft discrete\textquotedblright\ solitons in 2D photonic
lattices, Phys. Rev. Lett. \textbf{92}, 123904 (2004).

\bibitem{FPUT} J. Ford, The Fermi-Pasta-Ulam problem -- paradox turns
discovery, Phys. Rep. \textbf{213}, 271-310 (1992).

\bibitem{FK} O. M. Braun and Yu. S. Kivshar, \textit{The Frenkel-Kontorova
Model: Concepts, Methods, and Applications }(Springer, Berlin, 2004).

\bibitem{PT} Z. Chen, J. Liu, S. Fu, Y. Li, and B. A. Malomed, Discrete
solitons and vortices on two-dimensional lattices of $\mathcal{PT}$%
-symmetric couplers, Opt. Exp. \textbf{22}, 29679-29692 (2014).

\bibitem{Weinstein-2D} M. I. Weinstein, Excitation thresholds for nonlinear
localized modes on lattices, Nonlinearity \textbf{12}, 673-621 (1999).

\bibitem{Chong} C. Chong, R. Carretero-Gonz\'{a}lez, B. A. Malomed, and P.
G. Kevrekidis, Multistable solitons in higher-dimensional cubic-quintic
nonlinear Schr\"{o}dinger lattices, Physica D \textbf{238}, 126-136 (2009).

\bibitem{we} B. A. Malomed and P. G. Kevrekidis, Discrete vortex solitons,
Phys. Rev. E \textbf{64}, 026601 (2001).

\bibitem{they} J. Cuevas, G. James, P. G. Kevrekidis, and K. J. H. Law,
Vortex solutions of the discrete Gross-Pitaevskii equation starting from the
anti-continuum limit, Physica D \textbf{238}, 1422-1431 (2009).

\bibitem{Townes} R. Y. Chiao, E. Garmire, and C. H. Townes, Self-Trapping of
Optical Beams, Phys. Rev. Lett. \textbf{13}, 479-482 (1964).

\bibitem{Minsk} V. I. Kruglov, Yu. A. Logvin, and V. M. Volkov, The theory
of spiral laser beams in nonlinear media, J. Mod. Opt. \textbf{39},
2277-2291 (1992)

\bibitem{PhysD} B. A. Malomed, (INVITED)\ Vortex solitons: Old results and
new perspectives, Physica D \textbf{399}, 108-137 (2019).

\bibitem{Zhigang} P. G. Kevrekidis, B. A. Malomed, Z. Chen, and D. J.
Frantzeskakis, Stable higher-order vortices and quasivortices in the
discrete nonlinear Schr\"{o}dinger equation, Phys. Rev. E \textbf{70},
056612 (2004).

\bibitem{Zhig1} Z. Chen, M. Segev, D. W. Wilson, R. E. Muller, P. D. Maker,
Self-trapping of an optical vortex by use of the bulk photovoltaic effect,
Phys. Rev. Lett. \textbf{78}, 2948-2951 (1997).

\bibitem{Zhig2} Z. Chen, M.-F. Shih, M. Segev, D. W. Wilson, R. E. Muller,
and P. D. Maker, Steady-state vortex-screening solitons formed in biased
photorefractive media, Opt. Lett. \textbf{22}, 1751-1753 (1997).

\bibitem{Chen} A. Bezryadina, E. Eugenieva, and Z. Chen, Self-trapping and
flipping of double-charged vortices in optically induced photonic lattices,
Opt. Lett. \textbf{31}, 2456-2458 (2006).

\bibitem{Denz} B. Terhalle, T. Richter, K. J. H. Law, D. G\"{o}ries, P.
Rose, T. J. Alexander, P. G. Kevrekidis, Anton S. Desyatnikov, W.
Krolikowski, F. Kaiser, C. Denz, and Y. S. Kivshar, Observation of
double-charge discrete vortex solitons in hexagonal photonic lattices, Phys.
Rev. A \textbf{79}, 043821 (2009).

\bibitem{Sterke} C. M. de Sterke and J. E. Sipe, Gap solitons, Prog. Opt.,
vol. XXXIII, pp. 203-260 (1994).

\bibitem{Brazhnyi} V. A. Brazhnyi and V. V. Konotop, Theory of nonlinear
matter waves in optical lattices, Mod. Phys. Lett. B \textbf{18}, 627-651
(2004).

\bibitem{Morsch} O. Morsch and M. Oberthaler, Dynamics of Bose-Einstein
condensates in optical lattices, Rev. Mod. Phys. \textbf{78}, 179-212 (2006).

\bibitem{Oberthaler} B. Eiermann., Th. Anker, M. Albiez, M. Taglieber, P.
Treutlein, K.-P. Marzlin, and M. K. Oberthaler, Bright Bose-Einstein gap
solitons of atoms with repulsive interaction, Phys. Rev. Lett. \textbf{92},
230401 (2004).

\bibitem{Mok} J. T. Mok, C. M. de Sterke, I. C. M. Litte, and B. J.
Eggleton, Dispersionless slow light using gap solitons, Nature Phys. \textbf{%
2}, 775-780 (2006).

\bibitem{Belgrade} G. Gligori\'{c}, A. Maluckov, L. Hadzievski, and B. A.
Malomed, Localized modes in mini-gaps opened by periodically modulated
intersite coupling in two-dimensional nonlinear lattices, Chaos \textbf{24},
023124 (2014).

\bibitem{aniso} P. G. Kevrekidis, D. J. Frantzeskakis, R. Carretero-Gonz\'{a}%
lez, B. A. Malomed, and A. R. Bishop. Discrete solitons and vortices on
anisotropic lattices, Phys. Rev. E \textbf{72}, 046613 (2005).

\bibitem{JC2} J. Cuevas, B. A. Malomed, and P. G. Kevrekidis,
Two-dimensional discrete solitons in rotating lattices, Phys. Rev. E \textbf{%
76}, 046608 (2007).

\bibitem{potential} B. A. Malomed, Potential of interaction between two- and
three-dimensional solitons, Phys. Rev. E \textbf{58}, 7928-7933 (1998).

\bibitem{SSB} B. A. Malomed, Spontaneous symmetry breaking in nonlinear
systems: An overview and a simple model, in: \textit{Nonlinear Dynamics:
Materials, Theory and Experiments}, ed. by M. Tlidi, M. Clerc, Springer
Proceedings in Physics, vol. 173 (Springer, Cham, 2016), p. 97.

\bibitem{Iooss} G. Iooss, D. D. Joseph, \textit{Elementary Stability
Bifurcation Theory} (Springer, New York, 1980).

\bibitem{Cai} D. Cai, A. R. Bishop, and N. Gr{\o }nbech-Jensen, Perturbation
theories of a discrete, integrable nonlinear Schr\"{o}dinger equation, Phys.
Rev. E \textbf{53}, 4131-4136 (1996).

\bibitem{Cai97} D. Cai, A. R. Bishop, and N. Gr{\o }nbech-Jensen, Resonance
in the collision of two discrete intrinsic localized excitations, Phys. Rev.
E \textbf{56}, 7246-7252 (1997).

\bibitem{Dmitriev03} S. V. Dmitriev, P. G. Kevrekidis, B. A. Malomed, and D.
J. Frantzeskakis, Two-soliton collisions in a near-integrable lattice
system, Phys. Rev. E \textbf{68}, 056603 (2003).

\bibitem{Vakh} N. G. Vakhitov and A. A. Kolokolov, Stationary solutions of
the wave equation in a medium with nonlinearity saturation, Radiophys.
Quantum Electron. \textbf{16}, 783-789 (1973).

\bibitem{Zaragoza2D} J. G\'{o}mez-Garde\~{n}es, B. A. Malomed, L. M. Flor%
\'{\i}a, and A. R. Bishop, Discrete solitons and vortices in the
two-dimensional Salerno model with competing nonlinearities, Phys. Rev. E
\textbf{74}, 036607 (2006).

\bibitem{Spielman} Y.-J. Lin, K. Jim\'{e}nez-Garc\'{\i}a, and I. B.
Spielman, Spin-orbit-coupled Bose--Einstein condensates, Nature \textbf{471}%
, 83--86 (2011).

\bibitem{Galitski} V. Galitski and I. B. Spielman, Spin-orbit coupling in
quantum gases, Nature \textbf{494}, 49-54 (2013).

\bibitem{Goldman} N. Goldman, G. Juzeliunas, P. \"{O}hberg, and I. B.
Spielman, Light-induced gauge fields for ultracold atoms, Rep. Prog. Phys.
\textbf{77}, 126401 (2014).

\bibitem{Zhai} H. Zhai, Degenerate quantum gases with spin-orbit coupling,
Rep. Prog. Phys. \textbf{78}, 026001 (2015).

\bibitem{EPL} B. A. Malomed, Creating solitons by means of spin-orbit
coupling, EPL \textbf{122}, 36001 (2018).

\bibitem{Ben Li} H. Sakaguchi, B. Li, and B. A. Malomed, Creation of
two-dimensional composite solitons in spin-orbit-coupled self-attractive
Bose-Einstein condensates in free space, Phys. Rev. E \textbf{89}, 032920
(2014).

\bibitem{HS} H. Sakaguchi and B. A. Malomed, Discrete and continuum
composite solitons in Bose-Einstein condensates with the Rashba spin-orbit
coupling in one and two dimensions, Phys. Rev. E \textbf{90}, 062922 (2014).

\bibitem{3Dsol1} P. G. Kevrekidis, B. A. Malomed, D. J. Frantzeskakis and R.
Carretero-Gonz\'{a}lez, Three-dimensional solitary waves and vortices in a
discrete nonlinear Schr\"{o}dinger lattice, Phys. Rev. Lett. \textbf{93},
080403 (2004).

\bibitem{3Dsol2} R. Carretero-Gonz\'{a}lez, P. G. Kevrekidis, B. A. Malomed
and D. J. Frantzeskakis, Three-dimensional nonlinear lattices: From oblique
vortices and octupoles to discrete diamonds and vortex cubes, Phys. Rev.
Lett. \textbf{94}, 203901 (2005).

\bibitem{skyrm1} J. Ruostekoski and J. R. Anglin, Creating vortex rings and
three-dimensional skyrmions in Bose-Einstein condensates, Phys. Rev. Lett.
\textbf{86}, 3934-3937 (2001).

\bibitem{skyrm2} U. Al Khawaja and H. Stoof, Skyrmions in a ferromagnetic
Bose-Einstein condensate, Nature \textbf{411}, 918--920 (2001).

\bibitem{skyrm3} R. A. Battye, N. R. Cooper, and P. M. Sutcliffe, Stable
skyrmions in two-component Bose-Einstein condensates, Phys. Rev. Lett.
\textbf{88}, 080401 (2002).

\bibitem{skyrmion} P. G. Kevrekidis, R. Carretero-Gonz\'{a}lez, D. J.
Frantzeskakis, B. A. Malomed, and F. K. Diakonos. Skyrmion-like states in
two- and three-dimensional dynamical lattices. Phys. Rev. E \textbf{75},
026603 (2007).

\bibitem{baby1} A. Kudryavtsevy, B. Piette, and W. J. Zakrzewski, Skyrmions
and domain walls in (2+1) dimensions, Nonlinearity \textbf{11}, 783-795
(1998).

\bibitem{baby2} T. Weidig, The baby Skyrme models and their multi-skyrmions,
Nonlinearity \textbf{12}, 1489-1503 (1999).

\bibitem{Rubenchik} A. B. Aceves, C. De Angelis, A. M. Rubenchik, and S. K.
Turitsyn, Multidimensional solitons in fiber arrays, Opt. Lett. \textbf{19},
329 (1994).

\bibitem{Jena} S. Minardi, F. Eilenberger, Y. V. Kartashov, A. Szameit, U. R%
\"{o}pke, J. Kobelke, K. Schuster, H. Bartelt, S. Nolte, L. Torner, F.
Lederer, A. T\"{u}nnermann, and T. Pertsch, Three-dimensional light bullets
in arrays of waveguides, Phys. Rev. Lett. \textbf{105}, 263901 (2010).

\bibitem{Jena-vort} F. Eilenberger, K. Prater, S. Minardi, R. Geiss, U. R%
\"{o}pke, J. Kobelke, K. Schuster, H. Bartelt, S. Nolte, A. T\"{u}nnermann,
and T. Pertsch, Observation of discrete, vortex light bullets, Phys. Rev. X
\textbf{3}, 041031 (2013).

\bibitem{Sandra} P. Beli\v{c}ev, G. Gligori\'{c}, J. Petrovi\'{c}, A.
Maluckov, L. Had\v{z}ievski, B.A. Malomed, Composite localized modes in
discretized spin-orbit-coupled Bose-Einstein condensates, J. Phys. B\ At.
Mol. Opt. Phys. \textbf{48}, 065301 (2015).

\bibitem{Tamm} Yu. S. Kivshar, Nonlinear Tamm states and surface effects in
periodic photonic structures, Laser Phys. Lett. \textbf{5}, 703-713 (2008)

\bibitem{top-ins} D. R. Gulevich, D. Yudin, D. V. Skryabin, I. V. Iorsh, and
I. A. Shelykh, Edge solitons in kagome lattice, Sci. Rep. \textbf{7}, 1780
(2017).

\bibitem{top-ins2} Y. V. Kartashov and D. V. Skryabin, Modulational
instability and solitary waves in polariton topological insulators, Optica
\textbf{3}, 1228 (2016).

\bibitem{Hakim} V. Hakim and W. J. Rappel, Dynamics of the globally coupled
complex Ginzburg-Landau equation, Phys. Rev. A \textbf{46}, 7347-7350(R)
(1992).

\bibitem{Efremidis1} N. K. Efremidis and D. N. Christodoulides, Discrete
Ginzburg-Landau solitons, Phys. Rev. E \textbf{67}, 026606 (2003).

\bibitem{Akhmed} K. Maruno, A. Ankiewicz, and N. Akhmediev, Exact localized
and periodic solutions of the discrete complex Ginzburg-Landau equation,
Opt. Commun. \textbf{221}, 199-209 (2003).

\bibitem{Efremidis2} N. K. Efremidis, D. N. Christodoulides,\ and K.
Hizanidis, Two-dimensional discrete Ginzburg-Landau solitons, Phys. Rev. A
\textbf{76}, 043839 (2007).

\bibitem{Bender} C. M. Bender, Making sense of non-Hermitian Hamiltonians,
Rep. Prog. Phys. \textbf{70}, 947-1018 (2007).

\bibitem{Christod} K. G. Makris, R. El-Ganainy, D. N. Christodoulides, and
Z. H, Musslimani, Beam dynamics in $\mathcal{PT}$ symmetric optical
lattices, Phys. Rev. Lett. \textbf{100}, 103904 (2008).

\bibitem{Christod2} C. E. R\"{u}ter, K. G. Makris, R. El-Ganainy, D. N.
Christodoulides, M. Segev, and D. Kip, Observation of parity--time symmetry
in optics, Nature Phys. \textbf{6}, 192-195 (2010).

\bibitem{PTrev1} V. V. Konotop, J. Yang, and D. Zezyulin, Nonlinear waves in
$\mathcal{PT}$ -symmetric systems, Rev. Mod. Phys. \textbf{88}, 035002 (2016)

\bibitem{PTrev2} S. V. Suchkov, A. A. Sukhorukov, J. Huang, S. V. Dmitriev,
C. Lee, and Yu. S. Kivshar, Nonlinear switching and solitons in $\mathcal{PT}
$-symmetric photonic systems, Laser Phot. Rev. \textbf{10}, 177-213 (2016).

\bibitem{Radik} R. Driben and B. A. Malomed, Stability of solitons in
parity-time-symmetric couplers, Opt. Lett. \textbf{36}, 4323-4325 (2011).

\bibitem{Barash} N. V. Alexeeva , I. V. Barashenkov, A. A. Sukhorukov, and
Y. S. Kivshar, Optical solitons in $\mathcal{PT}$ -symmetric nonlinear
couplers with gain and loss, Phys. Rev. A \textbf{85}, 063837 (2012).

\bibitem{Gena} G. Burlak and B. A. Malomed, Stability boundary and
collisions of two-dimensional solitons in $\mathcal{PT}$-symmetric couplers
with the cubic-quintic nonlinearity, Phys. Rev. E \textbf{88}, 062904 (2013).

\bibitem{PTsol1} V. V. Konotop, D. E. Pelinovsky, and D. A. Zezyulin,
Discrete solitons in $\mathcal{PT}$-symmetric lattices, Europhys. Lett.
\textbf{100}, 56006 (2012).

\bibitem{PTsol0} C. Huang, C. Li, and L. Dong, Stabilization of
multipole-mode solitons in mixed linear-nonlinear lattices with a $\mathcal{%
PT}$ symmetry, Opt. Exp. \textbf{21}, 3917-3925 (2013).

\bibitem{PTsol2} D. Leykam, V. V. Konotop, and A. S. Desyatnikov, Discrete
vortex solitons and parity time symmetry, Opt. Lett. \textbf{38}, 371-373
(2013).

\bibitem{PTsol4} D. E. Pelinovsky, D. A. Zezyulin, and V. V. Konotop,
Nonlinear modes in a generalized $\mathcal{PT}$-symmetric discrete nonlinear
Schr\"{o}dinger equation, J. Phys. A: Math. Gen. \textbf{47}, 085204 (2014).

\bibitem{PTsol5} J. D'Ambroise, P. G. Kevrekidis, and B. A. Malomed,
Staggered parity-time-symmetric ladders with cubic nonlinearity, Phys. Rev.
E \textbf{91}, 033207 (2015).

\bibitem{PTsol-observation} M. Wimmer, A. Regensburger, M. A. Miri, C.
Bersch, D. N. Christodoulides, U. Peschel, Observation of optical solitons
in $\mathcal{PT}$-symmetric lattices, Nature Commun. \textbf{6}, 7782 (2015).
\end{thebibliography}
\end{document}